\documentclass[journal]{IEEEtran}

\usepackage{pgfplots}
\usepackage[bookmarks,colorlinks]{hyperref}
\usepackage[linesnumbered,ruled,lined]{algorithm2e}
\usepackage{enumitem}
\usetikzlibrary{shapes.multipart,intersections}
\usepackage{cite}
\usepackage{amsmath,amssymb,amsfonts,amsthm,steinmetz}
\usepackage{mathrsfs}  
\usepackage{textcomp}
\usepackage{acronym}
\usepackage{xcolor}
\usepackage{upgreek,xspace}
\usepackage{array}
\usepackage{tikz}
\usetikzlibrary{calc}
\makeatletter
\newcommand{\gettikzxy}[3]{%
  \tikz@scan@one@point\pgfutil@firstofone#1\relax
  \edef#2{\the\pgf@x}%
  \edef#3{\the\pgf@y}%
}
\makeatother

\usepackage{comment}

\usepackage[draft]{todonotes}   
\usepackage[T1]{fontenc}

\usepackage{esvect}

\usepackage{times}
\usepackage{bm}
\usepackage{stmaryrd}
\usepackage{babel}
\usepackage{graphics, graphicx}
\usepackage{gensymb}
\usepackage{cite}
\usepackage{enumitem}
\usepackage{url}

\begin{document}

\title{Electromagnetically Consistent Bounds on Information Transfer in \\Real-World RIS-Parametrized Wireless Channels}

\author{Albert~Salmi,~Ville~Viikari,~\IEEEmembership{Senior Member,~IEEE},~and~Philipp~del~Hougne,~\IEEEmembership{Member,~IEEE}
\thanks{
A.~Salmi, V.~Viikari, and P.~del~Hougne are with the Department of Electronics and Nanoengineering, Aalto University, 00076 Espoo, Finland. P. del Hougne is also with Univ Rennes, CNRS, IETR - UMR 6164, F-35000, Rennes, France. (e-mail: albert.salmi@aalto.fi; ville.viikari@aalto.fi; philipp.del-hougne@univ-rennes.fr)
}
\thanks{\textit{(Corresponding Author: Philipp del Hougne.)}}
\thanks{This work was supported in part by the Nokia Foundation (project 20260028), the ANR France 2030 program (project ANR-22-PEFT-0005), the ANR PRCI program (project ANR-22-CE93-0010), the Rennes M\'etropole AES program (project ``SRI''), the European Union's European Regional Development Fund, the French region of Brittany and Rennes Métropole through the contrats de plan État-Région program (projects ``SOPHIE/STIC \& Ondes'' and ``CyMoCoD''), and Business Finland (project MULTIRACS).}
}

\maketitle

\begin{abstract}
A reconfigurable intelligent surface (RIS) endows a wireless channel with programmability that can be leveraged to optimize wireless information transfer. While many works study algorithms for optimizing such a programmable channel, relatively little is known about fundamental bounds on the achievable information transfer. In particular, non-trivial bounds that are both electromagnetically consistent (e.g., aware of mutual coupling) and in line with realistic hardware constraints (e.g., few-bit-programmable, potentially lossy loads) are missing. Here, based on a rigorous multiport network model of a single-input single-output (SISO) channel parametrized by 1-bit-programmable RIS elements, we apply a semidefinite relaxation (SDR) to derive a fundamental bound on the achievable SISO channel gain enhancement. A bound on the maximum achievable rate of information transfer at a given noise level follows directly from Shannon's theorem. We apply our bound to several numerical and experimental examples of different RIS-parametrized radio environments. Compared to electromagnetically consistent benchmark bounding strategies (a norm-inequality bound and, where applicable, a relaxation to an idealized beyond-diagonal load network only constrained by losslessness and reciprocity for which a global solution exists), we consistently observe that our SDR-based bound is notably tighter. We reach at least 64\ \% (but often 100\ \%) of our SDR-based bound with standard discrete optimization techniques. The applicability of our bound to concrete experimental systems makes it valuable to inform wireless practitioners, e.g., to evaluate RIS hardware design choices and algorithms to optimize the RIS configuration. Our work contributes to the development of an electromagnetic information theory for RIS-parametrized channels as well as other programmable wave systems such as dynamic metasurface antennas or real-life beyond-diagonal RISs.

\end{abstract}


\section{Introduction}
\label{sec_introduction}

Shannon's theorem for the achievable rate of information transfer via a noisy channel underpins modern communications engineering and information theory~\cite{Shannon1948}. Specifically, Shannon’s channel coding theorem establishes the supremum of achievable reliable information rates, referred to as Shannon capacity, for a noisy single-input single-output (SISO) channel. The Shannon capacity $C$ (normalized by bandwidth) depends chiefly on the signal-to-noise ratio (SNR) with which the receiver detects the transmitted signal:
\begin{equation}
    C = \mathrm{log}_2 \left( 1 + \frac{P_\mathrm{T}}{\sigma^2} |h|^2 \right),
\end{equation}
where $P_\mathrm{T}$ denotes the transmit power, $\sigma^2$ quantifies the noise power, and $h$ denotes the channel gain, such that the term $\frac{P_\mathrm{T}}{\sigma^2} |h|^2$ is the SNR. 

The recent emergence of reconfigurable intelligent surfaces (RISs) has introduced the possibility of optimizing the channel gain. An RIS-parametrized channel is programmable, such that it becomes a function of the RIS control vector. We thus denote a programmable channel with $h(\mathbf{v})$, where $\mathbf{v}\in\mathcal{V}$ is the RIS control vector and $\mathcal{V}$ is its feasibility set; rigorous definitions are provided in Sec.~\ref{sec_SystemModel}. Importantly, a properly defined $h(\mathbf{v})$ captures electromagnetic interactions between RIS elements and the feasibility set $\mathcal{V}$ accounts for hardware constraints such as 1-bit-programmability (which is very common in existing RIS prototypes).
Consequently, the Shannon capacity of a programmable channel depends on the RIS control vector:
\begin{equation}
    C(\mathbf{v}) = \mathrm{log}_2 \left( 1 + \frac{P_\mathrm{T}}{\sigma^2} |h(\mathbf{v})|^2 \right),
    \label{eqCv}
\end{equation}
Inspection of (\ref{eqCv}) leads to a fundamental question: 

\textit{Shannon bounded the achievable information transfer in a fixed channel, but what is the bound on the achievable information transfer in a programmable channel subject to the constraints of an electromagnetically consistent $h(\mathbf{v})$ and a real-world feasibility set $\mathcal{V}$?} 

Given the monotone mapping from channel gain to Shannon capacity in (\ref{eqCv}), the question can be rephrased as 

\textit{What is the bound on the achievable channel gain in a programmable channel subject to the constraints of an electromagnetically consistent $h(\mathbf{v})$ and a real-world feasibility set $\mathcal{V}$?
}

Answering this question is non-trivial because $\mathcal V$ is discrete for real-life RIS prototypes, and because electromagnetic interactions between RIS elements constrain the set of realizable effective channels $\{h(\mathbf v):\mathbf v\in\mathcal V\}$ in a non-trivial manner. While many recent works propose algorithms to optimize $\mathbf v$ under various simplifying assumptions, these algorithms typically only find local optima. Indeed, there is generally no proof that these algorithms find the global optimum. The only case with an electromagnetically consistent global solution that we are aware of (see~\cite{wu2025beyond} and our Sec.~\ref{subsec_IBD}) requires a specific feasibility set making multiple idealized assumptions (see our discussion in Sec.~\ref{sec_discussion}). For real-world feasibility sets, no scalable technique for identifying a global solution (with a certificate) is known. Any algorithm finding a local optimum is thus confronted with the fundamental doubt whether there may be other algorithms capable of finding substantially better local optima. Besides assessing the performance of optimization algorithms, a tight bound on the achievable channel gain is also important for wireless practitioners to determine whether deploying an RIS is potentially of interest in a given use case before even developing any optimization algorithm. However, to the best of our knowledge, the problem of bounding the achievable gain of a programmable channel, and thus the achievable Shannon capacity, subject to electromagnetic and feasibility constraints, has not been studied in the existing literature. 

In this paper, we fill this gap. Our contributions are summarized as follows.
\begin{enumerate}
    \item We derive three electromagnetically consistent bounds on the achievable channel gain for a real-world feasibility set. \textit{First}, we derive a simple norm-inequality (NI) bound. \textit{Second}, we derive a bound based on a relaxation to an idealized beyond-diagonal (IBD) load network for which a global solution exists. \textit{Third}, we derive a bound based on a semidefinite relaxation (SDR)~\cite{luo2010semidefinite} of a quadratic formulation of the channel gain maximization subject to binary programmability constraints for each RIS element.
    \item We formalize three gauge freedoms in the system model parameters. We show that optimizing these gauge freedoms can tighten the NI bound (yielding the NIO bound), and that the IBD bound and the SDR bound are agnostic to these parameter ambiguities.
    \item We evaluate our bounds based on a full-wave numerical simulation of a SISO channel parametrized by 64 1-bit-programmable RIS elements. We consider three different sets of binary loads, including one involving only lossless loads that is compatible with the IBD bound. We probe the tightness of our bounds based on comparisons with four discrete optimizations of the RIS configuration (genetic algorithm, coordinate descent, exhaustive search (when feasible), projected SDR solution). We systematically study the dependence on the number of RIS elements.
    \item We evaluate our bounds for four distinct experimental SISO channels parametrized by 100 1-bit-programmable RIS elements. Again, we probe the tightness of our bounds based on comparisons with four discrete optimizations of the RIS configurations. We systematically study the dependence on the number of RIS elements.
\end{enumerate}

Our results reveal a clear hierarchy in terms of the tightness of the three bounds, with our NIO bound being the loosest one and our SDR bound being the tightest one. In fact, we observe a remarkable tightness of our SDR bound. Our discrete optimizations reach at least $64\ \%$ of our SDR bound. In three of our four experimental scenarios, our discrete optimizations reach our SDR bound (almost) exactly. Thereby, in these cases, we can certify that no optimization algorithms achieving substantially higher channel gains can exist. Given the exponential number of feasible RIS configurations ($2^{100}$ for 100 1-bit-programmable RIS elements), this is a remarkable conclusion.

Our present work contributes to the development of an \textit{electromagnetic information theory (EIT) for reconfigurable wave systems}~\cite{di2024electromagnetic}. Existing works on EIT are mostly focused on the effective electromagnetic degrees of freedom (EEMDOFs) in \textit{fixed} multiple-input multiple-output systems~\cite{franceschetti2009capacity,miller2019waves,shiu2000fading,miller2000communicating,piestun2000electromagnetic,poon2005degrees,migliore2006role,muharemovic2008antenna,yuan2021electromagnetic,pizzo2022landau,yuan2023effects,ruiz2023degrees}. Some recent works treat RISs as a transmitter capable of generating arbitrary aperture fields in the spirit of holographic MIMO~\cite{dardari_JSAC,decarli2021communication}. Only a few works in the realm of EIT treat RISs as programmable scattering objects that endow an otherwise fixed channel with programmability. Some works formulate EIT-based optimization objectives for RISs such as maximizing the number of EEMDOFs in an RIS-parametrized MIMO channel~\cite{del2019optimally,del2019optimized,do2022line,ruizWSA}. Recently,~\cite{del2025effective} introduced an electromagnetically consistent definition and evaluation of EEMDOFs in MIMO backscatter communications, where the information to be transmitted is encoded into the RIS configuration.
However, fundamental electromagnetic bounds on the tunability of \textit{programmable} channels remain unknown, even in the SISO case (with the aforementioned exception based on a global solution for a specific, idealized feasibility set). Consequently, fundamental electromagnetically consistent bounds on the achievable information transfer remain unknown, too. The present work provides those for the SISO case.

In the antenna community, a few works have considered the SDR of quadratic formulations of optimization problems for reactively loaded antenna arrays~\cite{corcoles2015reactively,salmi2025optimization}. However, these works neither considered real-life feasibility sets (e.g., 1-bit-programmable and potentially lossy loads as in our work), nor did they apply their bounds to experimental systems, nor did they relate them to (electromagnetic) information theory. 

Our work also connects to a growing research activity on electromagnetically consistent bounds on the performance of nanophotonic devices where SDR plays an important role~\cite{molesky2020hierarchical,kuang2020computational,gustafsson2020upper,liska2021fundamental,chao2022physical,shim2024fundamental,amaolo2024maximum,gertler2025many,virally2025many}. However, none of these works have applied their bounds to experimental systems, and most of these works consider the inverse design of static rather than tunable nanophotonic devices. Interestingly, the computational cost of SDR-based methods\footnote{Standard solvers have polynomial runtimes, with scalings between $n^{3.5}$ and $n^{4.5}$, where $n$ is the number of degrees of freedom~\cite{luo2010semidefinite,salmi2025optimization}.} can become a challenge given the large number of designable degrees of freedom in nanophotonic devices that are many wavelengths large; meanwhile, the real-life RIS parametrized radio environments we consider are also very large in terms of wavelengths but the computational complexity of evaluating our SDR-based bound is determined by the number of tunable elements which is typically on the order of 100 or 1000, such that the computational tractability poses no challenge to us.

\textit{Notation:}
$\mathbb{R}$, $\mathbb{C}$, and $\mathbb{B}\triangleq\{0,1\}$ denote the sets of real, complex, and binary numbers, respectively.
$\jmath\triangleq\sqrt{-1}$ denotes the imaginary unit.
$(\cdot)^*$ denotes elementwise complex conjugation.
$(\cdot)^\top$ and $(\cdot)^\dagger$ denote transpose and conjugate transpose, respectively.
$\Re\{\cdot\}$ denotes real part.
$\|\cdot\|_2$ denotes Euclidean norm for vectors and spectral norm for matrices.
$\mathbf{I}_a$ denotes the $a\times a$ identity matrix.
$\mathbf 0$ and $\mathbf 1$ denote the all-zeros and all-ones vectors/matrices of appropriate sizes.
$\mathbf{A}=\mathrm{diag}(\mathbf{a})$ denotes the diagonal matrix whose diagonal entries are given by the vector $\mathbf{a}$.
$[\mathbf{x}]_i$ denotes the $i$th entry of vector $\mathbf{x}$. $[\mathbf{A}]_{i,j}$ denotes the $(i,j)$ entry of matrix $\mathbf{A}$.
$\mathbf{x}_{\mathcal B}$ denotes the subvector of $\mathbf{x}$ selected by the index set $\mathcal B$.
$\left[\mathbf{A}\right]_{\mathcal{B}\mathcal{C}}$ denotes the submatrix (block) of $\mathbf{A}$ selected by row indices $\mathcal{B}$ and column indices $\mathcal{C}$.
$\mathrm{tr}(\cdot)$ and $\mathrm{vec}(\cdot)$ denote the trace and column-stacking vectorization operators, respectively.
$\mathbf{A}\succeq\mathbf 0$ and $\mathbf{A}\succ\mathbf 0$ denote positive semidefinite and positive definite matrices, respectively.

\section{System Model}
\label{sec_SystemModel}

In this section, we describe our electromagnetically consistent system model for an end-to-end SISO channel parametrized by $N_\mathrm{S}$ 1-bit-programmable RIS elements. 
The purpose of our system model is to map the RIS control vector $\mathbf{v}\in\mathbb{B}^{N_\mathrm{S}}$, whose $i$th binary entry determines the configuration of the $i$th 1-bit-programmable RIS element, to the resulting end-to-end wireless SISO channel $h(\mathbf{v}) \in \mathbb{C}$. Therefore, our system model involves two steps. \textit{First}, the RIS control vector $\mathbf{v}$ is encoded into the physical scattering properties of the RIS (summarized by the RIS load vector $\mathbf{r}\in\mathbb{C}^{N_\mathrm{S}}$ defined below). \textit{Second}, the RIS load vector is mapped to the end-to-end wireless channel using multiport network theory (MNT). Our system model thus requires an encoding function $f_\mathrm{enc}$ and an MNT function $f_\mathrm{MNT}$~\cite{ContRIS_LWC,largeRIS_TCOM,del2025experimental,pWDCperspective}:
\begin{subequations}\label{eq:enc-mnt}
\begin{align}
\mathbf{v} &\;\rightarrow\; \mathbf{r}(\mathbf{v}) = f_{\mathrm{enc}}(\mathbf{v}), \label{eq:enc-mnt:a}\\
\mathbf{r}(\mathbf{v}) &\;\rightarrow\; h(\mathbf{v}) = f_{\mathrm{MNT}}\!\big(\mathbf{r}(\mathbf{v})\big). \label{eq:enc-mnt:b}
\end{align}
\end{subequations}
Throughout this paper, we adopt a power-wave definition of $h$, in line with the channel measurements that can be realized experimentally using common instruments, such as a vector network analyzer (VNA). Specifically, we define $h$ as the ratio between the outgoing power wave at the receive port and the incoming power wave at the transmit port.

To formulate our system model, we begin by partitioning the radio environment into three entities: (i) the $N_\mathrm{A}$ ports of the antennas via which waves are injected and received (in our SISO case, $N_\mathrm{A}=2$); (ii) the $N_\mathrm{S}$ tunable lumped elements (each RIS element is parametrized by one tunable lumped element); (iii) the ensemble of all static scattering objects (including any environmental scattering as well as any structural scattering by antennas and RIS elements). Each tunable lumped element can be represented as an additional ``virtual'' port terminated by a tunable load. We do not assume any specifics about the environmental or structural scattering, other than that it is linear, time-invariant, and passive. Entity (iii) is characterized by an $N$-port scattering matrix $\mathbf{S}\in\mathbb{C}^{N\times N}$, where $N=N_\mathrm{A}+N_\mathrm{S}$ and we use the same reference impedance of $Z_0=50\ \Omega$ at all ports to define $\mathbf{S}$. For simplicity, we assume below that the signal generator connected to the port of the transmitting antenna and the signal detector connected to the port of the receiving antenna are matched to $Z_0$, as is commonly the case in practice. Meanwhile, entity (ii) is characterized by an $N_\mathrm{S}$-port diagonal scattering matrix $\mathbf{\Phi}(\mathbf{r}) = \mathrm{diag}(\mathbf{r}) \in\mathbb{C}^{N_\mathrm{S} \times N_\mathrm{S}}$, where $\mathbf{r}=\left[ \rho_1, \rho_2, \dots, \rho_{N_\mathrm{S}} \right]^\top$ and $\rho_i$ denotes the reflection coefficient of the load associated with the tunable lumped element of the $i$th RIS element. 

As discussed in~\cite{ContRIS_LWC,largeRIS_TCOM,del2025experimental,pWDCperspective}, the encoding function takes a particularly simple form in the case of identical 1-bit-programmable RIS elements:
\begin{equation}
    \mathbf{r}(\mathbf{v}) = f_\mathrm{enc}(\mathbf{v}) = \alpha \mathbf{1} + (\beta-\alpha) \mathbf{v},
    \label{eq2}
\end{equation}
where $\alpha\in\mathbb{C}$ and $\beta\in\mathbb{C}$ are the two possible reflection coefficients of the loads associated with the tunable lumped elements. 

Meanwhile, as discussed in~\cite{matteo_universal,ContRIS_LWC,largeRIS_TCOM,del2025experimental,pWDCperspective}, the MNT function mapping $\mathbf{r}(\mathbf{v})$ to $h(\mathbf{v})$ takes the following form:
\begin{equation}
h(\mathbf{r}) = f_\mathrm{MNT}(\mathbf{r}) = h_0 + \mathbf{a}^\top\,\bigl(\mathbf{I}_{N_\mathrm{S}}\,-\,\mathbf{\Phi}(\mathbf{r})\,\mathbf{\Gamma}\bigr)^{-1}\,\mathbf{\Phi}(\mathbf{r})\,\mathbf{b},
\label{eq3}
\end{equation}
where $h_0 = \mathbf{S}_\mathcal{RT}\in\mathbb{C}$, $\mathbf{a} = \mathbf{S}_\mathcal{RS}^\top\in\mathbb{C}^{N_\mathrm{S}}$, $\mathbf{\Gamma} = \mathbf{S}_\mathcal{SS}\in\mathbb{C}^{N_\mathrm{S}\times N_\mathrm{S}}$, and $\mathbf{b} = \mathbf{S}_\mathcal{ST}\in\mathbb{C}^{N_\mathrm{S}}$. Here, $\mathcal{T}$, $\mathcal{R}$, and $\mathcal{S}$ denote the sets containing the port indices associated with the transmitting port, the receiving port, and the RIS elements, respectively.

To summarize, the parameters involved in our system model can be compactly summarized as follows:
\begin{equation}
\boldsymbol{\theta}
\triangleq
\left[ \alpha, \, \beta, \, h_0, \, \mathbf{a}^\top, \, \mathbf{b}^\top, \, \operatorname{vec}(\mathbf\Gamma)^\top\right]^\top \in  \mathbb{C}^{3+2N_\mathrm{S}+N_\mathrm{S}^2}.
\label{eq:theta_def}
\end{equation}
$\alpha$ and $\beta$ are required by $f_\mathrm{enc}$ while the remaining parameters are required by $f_\mathrm{MNT}$. To emphasize the dependence of the channel on the model parameters, we write
\begin{equation}
h(\mathbf v;\boldsymbol{\theta}) \triangleq \hat f(\mathbf v;\boldsymbol{\theta}),
\label{eq:h_hatf}
\end{equation}
where $\hat f$ is the result of the combined action of $f_\mathrm{enc}$ followed by $f_\mathrm{MNT}$.

\section{Parameter Ambiguities and Gauge Freedoms}
\label{sec_Ambiguities}

Most works involving physics-consistent models of RIS-parametrized channels assume that all required parameters (i.e., $\alpha$, $\beta$, $h_0$, $\mathbf{a}$, $\mathbf{b}$, and $\mathbf{\Gamma}$) are known unambiguously. In theoretical works, the MNT parameters are either evaluated analytically (in sufficiently simple scenarios)~\cite{gradoni2021end} or extracted from a full-wave numerical simulation~\cite{tapie2023systematic,zheng2024mutual}. Meanwhile, theoretical works typically assume that the possible values of $\rho_i$ are only constrained by losslessness, without explicitly discussing the encoding function or considering few-bit-programmability constraints.

To work with a real-world RIS-parametrized wireless channel, it is not only important to account for the encoding function (as done in our system model) but also to ensure that the model parameters accurately describe the experimental reality. To minimize model-reality mismatch, the model parameters are ideally estimated experimentally. Indeed, on the one hand, the detailed RIS design may be unknown to the wireless practitioner (e.g., being proprietary) or subject to fabrication inaccuracies. Meanwhile, on the other hand, the exact geometry and material composition of environmental scattering objects is typically unknown. In addition, even with perfect knowledge of all details, the computational complexity of a full-wave numerical simulation is likely prohibitively costly. 

Unfortunately, however, the model parameters cannot be measured directly in experiments. The main reasons are that, \textit{first}, the ``virtual'' ports associated with RIS elements are typically not connectorized and, \textit{second}, in any case, typical instruments like VNAs only have a limited number of ports (e.g., two, four, or eight) that is far below the order of magnitude of $N_\mathrm{S}$ for a large RIS (e.g., 64 or 100). Moreover, the parameters cannot be estimated unambiguously based on measurements of $h(\mathbf{v})$ for known $\mathbf{v}$. Indeed, an ambiguous experimental estimation of the MNT parameters would require knowledge of the encoding parameters as well as the availability of coupled loads, as detailed in recent works on a ``Virtual VNA'' technique~\cite{del2024virtual2p0,del2025wireless}. Nonetheless, an operationally equivalent proxy set of model parameters can be estimated experimentally. Specifically,~\cite{sol2024experimentally,del2025physics,ContRIS_LWC,largeRIS_TCOM,del2025experimental} report various techniques for estimating a set of proxy parameters (denoted with a tilde),
\begin{equation}
\tilde{\boldsymbol{\theta}}
\triangleq
\left[ \tilde{\alpha}, \, \tilde{\beta}, \, \tilde{h}_0, \, \tilde{\mathbf{a}}^\top, \, \tilde{\mathbf{b}}^\top, \, \operatorname{vec}(\tilde{\mathbf\Gamma})^\top\right]^\top \in  \mathbb{C}^{3+2N_\mathrm{S}+N_\mathrm{S}^2},
\label{eq:theta_deftilde}
\end{equation}
that can be used in lieu of the true parameters to accurately map any $\mathbf{v}$ to the corresponding $h(\mathbf{v})$ based on (\ref{eq2}) and (\ref{eq3}). In other words, the proxy parameters' operational equivalence refers to
\begin{equation}
    \hat{f}(\mathbf{v};\boldsymbol{\theta}) = \hat{f}(\mathbf{v};\tilde{\boldsymbol{\theta}}) \ \forall \ \mathbf{v}\in\mathbb{B}^{N_\mathrm{S}}.
\label{eq:opeq}
\end{equation}

In this section, we elucidate three gauge transformations of the true model parameters $\boldsymbol{\theta}$ to a valid set of proxy parameters $\tilde{\boldsymbol{\theta}}$. 
For the present paper, a clear understanding of these gauge transformations is important in order to be able to assess whether the computed bounds depend on the ambiguities (which are inevitable for experimentally estimated parameters), and, if they do, how to identify a gauge-agnostic bound.
Even if $\boldsymbol{\theta}$ is known, understanding these ambiguities is important because some gauge transformations may be able to tighten certain bounds, as seen below.

In the following, we describe each of the three gauge transformations in turn; they can be freely combined.

\textit{Diagonal-Similarity Gauge:}
Given $N_\mathrm{S}$ non-zero, complex-valued gauge parameters $d_i$, we define 
\begin{equation}
\label{eq:gauge_diag}
\begin{alignedat}{3}
\tilde h_0 &= h_0, &\qquad \tilde\alpha &= \alpha, &\qquad \tilde\beta &= \beta,\\
\tilde{\mathbf a}^\top &= \mathbf a^\top \mathbf D^{-1}, &\qquad
\tilde{\mathbf b} &= \mathbf D\,\mathbf b, &\qquad
\tilde{\mathbf\Gamma} &= \mathbf D\,\mathbf\Gamma\,\mathbf D^{-1},
\end{alignedat}
\end{equation}
where $\mathbf D=\mathrm{diag}(\mathbf{d})\in\mathbb{C}^{N_\mathrm{S}\times N_\mathrm{S}}$ and $\mathbf{d} = \left[d_1,\dots,d_{N_\mathrm S}\right]^\top\in\mathbb{C}^{N_\mathrm{S}}$.
This diagonal-similarity gauge can be interpreted as a per-port renormalization at the ``virtual'' ports associated with the RIS elements. While this diagonal-similarity gauge does not influence the load characteristics, it generally does not preserve a possible passivity or reciprocity of $\mathbf{\Gamma}$; $\tilde{\mathbf{\Gamma}}$ is symmetric only if $\mathbf{D}^2$ commutes with $\mathbf{\Gamma}$, e.g., when all $d_i$ are equal. 
To eliminate the ambiguity associated with this diagonal-similarity gauge, the ``Virtual VNA'' technique leverages coupled loads~\cite{del2024virtual2p0,del2025virtual3p0,del2025wireless}.

\textit{Complex-Scaling Gauge:}
Given a non-zero, complex-valued gauge parameter $c$, we define
\begin{equation}
\label{eq:gauge_scale}
\begin{alignedat}{3}
\tilde h_0 &= h_0, &\qquad \tilde\alpha &= c\,\alpha, &\qquad \tilde\beta &= c\,\beta,\\
\tilde{\mathbf a} &= \frac{1}{c}\,\mathbf a, &\qquad
\tilde{\mathbf b} &= \mathbf b, &\qquad
\tilde{\mathbf\Gamma} &= \frac{1}{c}\,\mathbf\Gamma.
\end{alignedat}
\end{equation}
This complex-scaling gauge influences the load characteristics; it preserves a possible reciprocity of $\mathbf{\Gamma}$ but not a possible passivity.
The ``Virtual VNA'' technique is not confronted with this complex-scaling gauge because the known loads eliminate this ambiguity.

\textit{M\"obius Gauge:}
Given a complex-valued gauge parameter $m$ and the M\"obius transformation $\mathcal M_m(\rho)\triangleq (\rho-m)/(1-m^*\rho)$, we define
\begin{equation}
\label{eq:gauge_mobius}
\begin{alignedat}{3}
\tilde h_0 &= h_0 + m\,\mathbf a^\top \mathbf F \,\mathbf b, &\; \; \;
\tilde\alpha &= \mathcal M_m(\alpha), &\; \; \; 
\tilde\beta &= \mathcal M_m(\beta),\\
\tilde{\mathbf a}^\top &= k\,\mathbf a^\top \mathbf F, &\;
\tilde{\mathbf b} &= k\,\mathbf F\,\mathbf b, &\;
\tilde{\mathbf\Gamma} &= (\mathbf\Gamma-m^*\mathbf I_{N_\mathrm{S}})\,\mathbf F,
\end{alignedat}
\end{equation}
where $\mathbf F \triangleq (\mathbf I_{N_\mathrm{S}}-m\mathbf\Gamma)^{-1}$ and $k\triangleq\sqrt{1-|m|^2}$. This M\"obius gauge corresponds to changing the scattering reference at the RIS virtual ports by absorbing a uniform baseline reflection into the load description. For numerical stability, it is necessary to enforce $m\neq 1/\rho^*$ for $\rho\in\{\alpha,\beta\}$ and a reasonable conditioning of $(\mathbf I_{N_\mathrm{S}}-m\mathbf\Gamma)$.
If $|m|<1$, a potential passivity of $\mathbf{\Gamma}$ is preserved. 
Three distinct and known terminations of the ``virtual'' ports associated with the RIS elements are required to eliminate this M\"obius ambiguity~\cite{del2024virtual2p0,del2025virtual3p0,ContRIS_LWC,del2025wireless}.

To summarize, the gauge parameters involved in the three described ambiguities can be compactly summarized as follows:
\begin{equation}
\boldsymbol{\phi}
\triangleq
\left[ \mathbf{d}^\top, \, c, \, m \right]^\top \in  \mathbb{C}^{2+N_\mathrm{S}}.
\label{eq:phi_def}
\end{equation}
We denote a transformation involving all three ambiguities as
\begin{equation}
    \tilde{\boldsymbol{\theta}} = \hat{g}({\boldsymbol{\theta}};\boldsymbol{\phi}),
\end{equation}
where we assume that $\hat{g}$ applies the gauges in the order they were presented in this section.

\section{Bounds on Achievable SISO Channel Gain}
\label{sec_Bounds}

In this section, we describe three bounds on the achievable SISO channel gain in a wireless channel parametrized by 1-bit-programmable RIS elements. The fundamental challenge is not merely to formulate a valid bound, but to formulate a valid bound that is tight because it takes practical details into account. Indeed, the fundamental consideration of passivity of the RIS-parametrized radio environment yields a trivial bound on the SISO channel gain: $|h(\mathbf{v})|^2 < 1 \ \forall \ \mathbf{v}\in\mathbb{B}^{N_\mathrm{S}}$. However, while valid, this trivial bound is generally very loose and of limited practical value as it does not incorporate sufficient practical details to be useful for guiding hardware or algorithmic design. Of the three bounds described in this section, the SDR-based bound is the most elaborate one, and we show that it is the tightest bound for the examples studied in subsequent sections. The norm-inequality bound is a relatively simple bound that serves as benchmark for the tightness of our SDR-based bound. The idealized-BD-RIS bound only applies if $|\alpha|=|\beta|=1$, as explained below, in which case it serves as an additional benchmark.

\subsection{Norm-Inequality (NI) Bound}
\label{subsec_NIB}

By applying the triangle inequality and the Cauchy-Schwarz inequality to (\ref{eq3}), we obtain
\begin{equation}
    |h(\mathbf{r})| \le |h_0| + \|\mathbf a\|_2\, \bigl\|(\mathbf{I}_{N_\mathrm{S}}-\mathbf{\Phi}(\mathbf{r})\,\mathbf\Gamma)^{-1}\bigr\|_2\, \|\mathbf\Phi(\mathbf{r})\|_2\, \|\mathbf b\|_2.
    \label{eq13}
\end{equation}
Since the spectral norm of a diagonal matrix equals the maximum modulus of its diagonal entries, 
\begin{equation}
\|\mathbf\Phi(\mathbf r)\|_2 = \max_i |\rho_i| \le \gamma,
\label{eq:NIB_phi_norm}
\end{equation}
where $\gamma \triangleq \max\{|\alpha|,|\beta|\}$. To upper-bound the resolvent term, we use the submultiplicativity of the spectral norm:
\begin{equation}
\|\mathbf\Phi(\mathbf{r})\,\mathbf\Gamma\|_2 \le \|\mathbf\Phi(\mathbf{r})\|_2\,\|\mathbf\Gamma\|_2 \le \gamma\|\mathbf\Gamma\|_2.
\label{eq:NIB_submult}
\end{equation}
If $\gamma\|\mathbf\Gamma\|_2<1$, then $\|\mathbf\Phi(\mathbf{r})\,\mathbf\Gamma\|_2<1$ and the Neumann series
$(\mathbf I_{N_\mathrm{S}}-\mathbf\Phi(\mathbf{r})\,\mathbf\Gamma)^{-1}=\sum_{k=0}^\infty (\mathbf\Phi(\mathbf{r})\,\mathbf\Gamma)^k$
converges. Then,
\begin{equation}\label{eq:NIB_resolvent}
\begin{aligned}
\bigl\|(\mathbf I_{N_\mathrm{S}}-\mathbf\Phi(\mathbf{r})\,\mathbf\Gamma)^{-1}\bigr\|_2
\le \sum_{k=0}^\infty \|\mathbf\Phi(\mathbf{r})\,\mathbf\Gamma\|_2^k
&= \frac{1}{1-\|\mathbf\Phi(\mathbf{r})\,\mathbf\Gamma\|_2}, \\
&\le \frac{1}{1-\gamma\|\mathbf\Gamma\|_2}.
\end{aligned}
\end{equation}
Combining \eqref{eq:NIB_phi_norm} and \eqref{eq:NIB_resolvent} with (\ref{eq13}) yields
\begin{equation}
|h(\mathbf r)|
\le |h_0| + \|\mathbf a\|_2\,\frac{\gamma}{1-\gamma\|\mathbf\Gamma\|_2}\,\|\mathbf b\|_2.
\label{eq:NIB_h_bound}
\end{equation}
Squaring both sides of (\ref{eq:NIB_h_bound}) yields the norm-inequality bound on the SISO channel gain:
\begin{equation}
|h(\mathbf r)|^2
\le B_\mathrm{NI} \triangleq 
\left(|h_0| + \|\mathbf a\|_2\,\frac{\gamma}{1-\gamma\|\mathbf\Gamma\|_2}\,\|\mathbf b\|_2\right)^2,
\label{eq:NIB_final}
\end{equation}
where a sufficient condition for validity is $\gamma\|\mathbf\Gamma\|_2<1$.

Because $B_\mathrm{NI}$ is computed based on norms of model-internal variables, it is generally \textit{not} agnostic to the ambiguities discussed in Sec.~\ref{sec_Ambiguities}. In fact, only the complex-scaling ambiguity does not impact $B_\mathrm{NI}$. The diagonal-similarity ambiguity generally impacts $B_\mathrm{NI}$, unless $|d_i|=1 \ \forall \ i$. The dependence of $B_\mathrm{NI}$ on $d_i$ and $m$ points to an opportunity for tightening the bound. Indeed, one can optimize the diagonal-similarity gauge parameters and the M\"obius gauge parameter in order to minimize $B_\mathrm{NI}$ and thereby tighten the bound:
\begin{equation}
\label{eq:gauge_opt_BNI}
\begin{aligned}
B_\mathrm{NIO}(\boldsymbol\theta)
\triangleq\ 
\min_{\mathbf d,\;m}\quad 
& B_\mathrm{NI}\!\left(\hat g(\boldsymbol\theta;[\mathbf d^\top,\,c_0,\,m]^\top)\right)\\
\text{s.t.}\quad 
& \tilde\gamma\,\bigl\|\tilde{\mathbf\Gamma}\bigr\|_2 < 1,\\
& d_i\neq 0\ \forall \ i,\\
& \text{all inverses in }\hat g(\cdot)\text{ exist},
\end{aligned}
\end{equation}
where $\tilde{\boldsymbol\theta}=\hat g(\boldsymbol\theta;[\mathbf d^\top,\,c_0,\,m]^\top)$ with any fixed $c_0\neq 0$ (e.g., $c_0=1$ without loss of optimality since $B_\mathrm{NI}$ is invariant under the complex-scaling gauge), and $\tilde\gamma\triangleq\max\{|\tilde\alpha|,|\tilde\beta|\}$. The problem in (\ref{eq:gauge_opt_BNI}) involves $1+N_\mathrm{S}$ complex-valued unknowns and can be solved with standard gradient-descent techniques, e.g., by automatic differentiation in frameworks such as TensorFlow or PyTorch.

\subsection{Idealized-BD-RIS (IBD) Bound}
\label{subsec_IBD}

If $|\alpha|=|\beta|=1$, then $\mathbf{\Phi}(\mathbf r)$ is diagonal with unit-modulus entries and hence unitary. In this particular case, our feasible set of $\mathbf\Phi(\mathbf r)$ is a strict subset of the set of all \emph{unitary} matrices. Consequently, an upper bound obtained by allowing $\mathbf\Phi$ to be \emph{any} unitary matrix also upper-bounds the case of a conventional RIS with 1-bit-programmable elements and $|\alpha|=|\beta|=1$.
A RIS with non-diagonal $\mathbf{\Phi}$ is commonly referred to as ``beyond-diagonal'' RIS (BD-RIS); if $\mathbf{\Phi}$ is not symmetric, the load network of the BD-RIS is non-reciprocal. Recently,~\cite{wu2025beyond} identified a technique for determining the globally optimal $\mathbf{\Phi}$ for SISO channel gain maximization with an idealized BD-RIS for which $\mathbf{\Phi}$ can be any unitary and reciprocal load network. Importantly, while the assumptions in~\cite{wu2025beyond} about the realizable $\mathbf{\Phi}$ are idealized, no simplifications of the MNT model (such as the unilateral approximation, or ignoring mutual coupling) were made in~\cite{wu2025beyond}. However, the definition of $h$ used in~\cite{wu2025beyond} is based on voltages rather than power waves. In this subsection, we adapt the technique presented in~\cite{wu2025beyond} so that we can determine the globally optimal solution for SISO channel gain maximization with the power-wave-based channel definition used in the present work. The resulting global solution for a BD-RIS with a lossless load network then constitutes an upper bound for the case of a conventional RIS with lossless loads. We omit demonstrating that the global solution can be attained by a load network that is not only unitary but also reciprocal because such a demonstration does not affect the value of the bound that is of interest in the present paper.

We consider the following optimization problem:
\begin{equation}
\max_{\mathbf\Phi\in\mathbb C^{N_\mathrm S\times N_\mathrm S}}\quad |h(\mathbf\Phi)|^2
\quad \text{s.t.}\quad  \mathbf\Phi^\dagger\mathbf\Phi=\mathbf I_{N_\mathrm S},
\label{eq:bd_opt}
\end{equation}
with the same multiport-network channel model as in \eqref{eq3}, except that $\mathbf\Phi$ is no longer constrained to be diagonal. We define the auxiliary variable
\begin{equation}
\mathbf x \triangleq \bigl(\mathbf{I}_{N_\mathrm{S}} -\mathbf\Phi\,\mathbf\Gamma\bigr)^{-1}\mathbf\Phi\,\mathbf b \in \mathbb{C}^{N_\mathrm{S}},
\label{eq:bd_x_def}
\end{equation}
so that 
\begin{equation}
h(\mathbf\Phi)=h_0+\mathbf a^\top\mathbf x.
\label{eq:eqn_w_x}
\end{equation}
Upon rearranging (\ref{eq:bd_x_def}) and using the unitarity of $\mathbf{\Phi}$, we find
\begin{equation}
    \mathbf{\Phi}^\dagger \, \mathbf{x} = \mathbf{b} + \mathbf{\Gamma}\,\mathbf{x}.
\end{equation}
The unitarity of $\mathbf{\Phi}$ further implies $\|\mathbf\Phi^\dagger \, \mathbf x\|_2=\|\mathbf x\|_2$, yielding
\begin{equation}
\|\mathbf x\|_2^2=\|\mathbf b+\mathbf\Gamma\,\mathbf x\|_2^2.
\label{eq:bd_energy_balance}
\end{equation}
Expanding \eqref{eq:bd_energy_balance} leads to
\begin{equation}
\begin{aligned}
p^2 \triangleq (\mathbf x-\mathbf x_0)^\dagger\,\mathbf Q\,(\mathbf x-\mathbf x_0)
&= \mathbf b^\dagger\,\mathbf b+\mathbf x_0^\dagger\,\mathbf Q\,\mathbf x_0 \\
&= \mathbf b^\dagger\,\mathbf b+\mathbf b^\dagger\,\mathbf\Gamma\,\mathbf Q^{-1}\mathbf\Gamma^\dagger\,\mathbf b,
\end{aligned}
\label{eq:bd_square_complete}
\end{equation}
where $\mathbf Q\triangleq \mathbf I_{N_\mathrm S}-\mathbf\Gamma^\dagger\,\mathbf\Gamma$ and $\mathbf x_0\triangleq \mathbf Q^{-1}\,\mathbf\Gamma^\dagger\,\mathbf b$.

Feasible vectors $\mathbf x$ satisfy the quadratic constraint given in~\eqref{eq:bd_square_complete}. $\mathbf{Q}$ can be interpreted as the unitarity deficit of the scattering matrix block $\mathbf{\Gamma}$ seen at the ``virtual'' ports associated with the RIS elements when the antenna ports are terminated with matched loads. The radio environment being passive and containing numerous loss and leakage mechanisms (including, but not limited to, the two matched loads at the antenna ports) implies the contractivity condition $\mathbf\Gamma^\dagger\,\mathbf\Gamma\prec \mathbf I_{N_\mathrm S}$ that yields $\mathbf Q\succ \mathbf 0$.
Geometrically, the quadratic constraint describes the surface of an ellipsoid in $\mathbb C^{N_\mathrm S}$ (centered at $\mathbf x_0$ with shape matrix $\mathbf Q$).
Since $|h(\mathbf\Phi)|^2=|h_0+\mathbf a^\top\mathbf x|^2$ is maximized on the ellipsoid surface, we may equivalently work with the filled ellipsoid
\begin{equation}
(\mathbf x-\mathbf x_0)^\dagger\,\mathbf Q\,(\mathbf x-\mathbf x_0)\le p^2.
\label{eq:bd_ellipsoid_filled}
\end{equation}
With $h_\mathrm c \triangleq h_0+\mathbf a^\top\mathbf x_0$ and 
$\mathbf y\triangleq \mathbf Q^{1/2}(\mathbf x-\mathbf x_0)$, where $\|\mathbf y\|_2\le p$, it follows that
\begin{equation}
h_0+\mathbf a^\top\mathbf x
=
h_\mathrm c + (\mathbf Q^{-1/2}\mathbf a^*)^\dagger\,\mathbf y.
\end{equation}
By the Cauchy--Schwarz inequality, the maximum is
\begin{equation}
\begin{aligned}
\max_{\mathbf\Phi^\dagger\mathbf\Phi=\mathbf I}\ |h(\mathbf\Phi)|
&= |h_\mathrm c| + p\,\|\mathbf Q^{-1/2}\mathbf a^*\|_2 \\
&= |h_\mathrm c| + p\,\sqrt{\mathbf a^\top\mathbf Q^{-1}\mathbf a^*}.
\end{aligned}
\label{eq:bd_bound_abs}
\end{equation}
We define the corresponding idealized-BD-RIS bound on the SISO channel gain as
\begin{equation}
B_\mathrm{IBD}\triangleq
\left(|h_\mathrm c| + p\,\sqrt{\mathbf a^\top\mathbf Q^{-1}\mathbf a^*}\right)^2.
\label{eq:bd_bound_final}
\end{equation}

What is left to prove is that the bound in (\ref{eq:bd_bound_final}) is achievable under the unitary constraint in (\ref{eq:bd_opt}). The choice $\mathring{\mathbf y}\triangleq \bigl(p/\|\mathbf Q^{-1/2}\mathbf a^*\|_2\bigr)\,e^{\jmath\arg(h_\mathrm c)}\,\mathbf Q^{-1/2}\mathbf a^*$ attains the bound in (\ref{eq:bd_bound_abs}). From the definition of $\mathring{\mathbf y}$ it follows that the corresponding $\mathring{\mathbf x}=\mathbf x_0+\mathbf Q^{-1/2}\mathring{\mathbf y}$ satisfies $(\mathring{\mathbf x}-\mathbf x_0)^\dagger\mathbf Q(\mathring{\mathbf x}-\mathbf x_0)=p^2$. Returning from (\ref{eq:bd_square_complete}) to (\ref{eq:bd_energy_balance}), it follows that $\|\mathring{\mathbf x}\|_2=\|\mathbf b+\mathbf\Gamma\mathring{\mathbf x}\|_2$. Thus, there exists a unitary matrix $\mathring{\mathbf\Phi}$ such that $\mathring{\mathbf\Phi}^\dagger\mathring{\mathbf x}=\mathbf b+\mathbf\Gamma\mathring{\mathbf x}$. Hence, $\mathring{\mathbf x}$ is realizable under the unitary constraint in (\ref{eq:bd_opt}) and attains the value in (\ref{eq:bd_bound_abs}); therefore the bound in (\ref{eq:bd_bound_final}) is achievable.

To derive the bound in (\ref{eq:bd_bound_final}) and its achievability, we only used the unitarity of $\mathbf{\Phi}$. In this paper, we do not demonstrate that this bound is achievable with a unitary \textit{and symmetric} $\mathbf{\Phi}$ because such a demonstration would not affect the value of the bound that is of interest here.\footnote{Readers interested in demonstrating the feasibility of achieving the globally optimal solution with a lossless \textit{and reciprocal} load network may refer to Algorithm~1 in~\cite{nerini_closedform_noMC} which solves a related problem.}

Importantly, we can only apply this bound when $|\alpha| = |\beta|=1$ \textit{and} $\mathbf\Gamma^\dagger\,\mathbf\Gamma\prec \mathbf I_{N_\mathrm S}$; the latter is required to ensure $\mathbf Q\succ \mathbf 0$ in \eqref{eq:bd_square_complete}. Since $B_\mathrm{IBD}$ is achievable under the idealized assumptions of a lossless BD-RIS load network, it represents the globally optimal value of \eqref{eq:bd_opt} and is therefore necessarily invariant to reparametrizations with gauge parameters that keep the requirements $|\alpha|=|\beta|=1$ and $\mathbf\Gamma^\dagger\mathbf\Gamma\prec \mathbf I_{N_\mathrm S}$ satisfied. In particular, unlike the norm-inequality bound, $B_\mathrm{IBD}$ thus cannot be tightened by optimizing gauge parameters.

A distinct question is whether, given a set of model parameters that does \emph{not} satisfy $|\alpha|=|\beta|=1$ \textit{and} $\mathbf\Gamma^\dagger\,\mathbf\Gamma\prec \mathbf I_{N_\mathrm S}$, there exists a choice of gauge parameters $\boldsymbol\phi$ such that the transformed parameters $\tilde{\boldsymbol\theta}=\hat g(\boldsymbol\theta;\boldsymbol\phi)$ satisfy $|\tilde\alpha|=|\tilde\beta|=1$ and $\tilde{\mathbf\Gamma}^\dagger\,\tilde{\mathbf\Gamma}\prec \mathbf I_{N_\mathrm S}$, thereby enabling the use of the idealized-BD-RIS bound. It is not clear that such a reparametrization is necessarily possible.

\subsection{SDR-Based (SDR) Bound}
\label{sec:sdr_bound}

Based on (\ref{eq:eqn_w_x}), we can express the SISO channel gain (i.e., the quantity that we wish to maximize) as  
\begin{equation}
\begin{aligned}
|h(\mathbf{x})|^2
&= \bigl(h_0+\mathbf a^\top \mathbf x\bigr)\bigl(h_0^*+(\mathbf a^\top \mathbf x)^*\bigr) \\
&= \mathbf{x}^\dagger\,\mathbf{R}_0\,\mathbf{x}
+ 2\,\Re\!\left\{\mathbf{q}_0^\top\,\mathbf{x}\right\}
+ t_0 .
\end{aligned}
\label{eq:gain_quad}
\end{equation}
where $\mathbf{R}_0 = \mathbf{a}^*\mathbf{a}^\top$, $\mathbf{q}_0^\top = h_0^*\,\mathbf{a}^\top$, and $t_0 = |h_0|^2$. Our objective to maximize the SISO channel gain is thus quadratic in $\mathbf{x}$. 
The formulation in (\ref{eq:gain_quad}) based on the auxiliary variable $\mathbf{x}$ parallels the one used to study a reactively loaded antenna array in~\cite{salmi2025optimization}.

Besides expressing our objective in quadratic form, we aim to account for the 1-bit-programmability constraint of our RIS elements in quadratic form. To formulate the binary-programmability constraint for the $i$th RIS element in quadratic form, we note (inspired by conceptually related work in theoretical nanophotonics~\cite{shim2024fundamental,gertler2025many}) that it can be interpreted as a logical OR condition and written as $(\rho_i-\alpha)(\rho_i-\beta)=0$. It is obvious that the condition
\begin{equation}
 (\rho_i-\alpha)^*(\rho_i-\beta)=0 
 \label{eq_binary_rho}
\end{equation}
also holds. To express this binary-programmability constraint from (\ref{eq_binary_rho}) in terms of our auxiliary variable $\mathbf{x}$, we note that the definition of $\mathbf{x}$ in (\ref{eq:bd_x_def}) implies $(\mathbf{I}_{N_\mathrm{S}}-\mathbf{\Phi}\,\mathbf{\Gamma})\,\mathbf{x} = \mathbf{\Phi}\,\mathbf{b}$ which leads to $\mathbf{x} = \mathbf{\Phi}(\mathbf{b}+\mathbf{\Gamma}\,\mathbf{x})$. With $\mathbf z\triangleq \mathbf b+\mathbf\Gamma\mathbf x$, this identity is equivalent to the elementwise relations $x_i=\rho_i z_i$ for all $i\in\{1,\dots,N_\mathrm S\}$, where $z_i=b_i+(\mathbf\Gamma\mathbf x)_i$.
Because $\rho_i$ equals either $\alpha$ or $\beta$, $x_i$ equals either $\alpha z_i$ or $\beta z_i$. Using the same reasoning that led to (\ref{eq_binary_rho}), we thus have
\begin{equation}
(x_i-\alpha z_i)^*(x_i-\beta z_i)=0.
\label{eq32}
\end{equation}
The binary programmability constraint for the $i$th RIS element can thus be expressed as a quadratic equality in $\mathbf x$. To obtain a quadratic form of~(\ref{eq32}) in the vector variable $\mathbf{x}$, we note that $x_i=\mathbf u_i^\top\mathbf x$, where $\mathbf u_i\in\mathbb C^{N_\mathrm S}$ denotes the $i$th canonical basis vector, and $z_i=b_i+\mathbf c_i^\top\mathbf x$, where $\mathbf c_i\in\mathbb C^{N_\mathrm S}$ and $\mathbf c_i^\top$ is the $i$th row of $\mathbf\Gamma$. We can thus rewrite~(\ref{eq32}) as 
\begin{equation}
\bigl(\mathbf g_{\alpha,i}^\top\mathbf x-\alpha b_i\bigr)^*
\bigl(\mathbf g_{\beta,i}^\top\mathbf x-\beta b_i\bigr)=0,
\label{eq33}
\end{equation}
where $\mathbf g_{\alpha,i}^\top \triangleq \mathbf u_i^\top-\alpha\,\mathbf c_i^\top$ and $\mathbf g_{\beta,i}^\top \triangleq \mathbf u_i^\top-\beta\,\mathbf c_i^\top$. Then, we can expand (\ref{eq33}) as follows:
\begin{equation}
\mathbf x^\dagger \mathbf R_i \mathbf x +\mathbf x^\dagger \mathbf q_{1,i} +\mathbf q_{2,i}^\top \mathbf x +t_i = 0,
\label{eq:bin_qcqp}
\end{equation}
where $\mathbf R_i \triangleq \mathbf g_{\alpha,i}^*\,\mathbf g_{\beta,i}^\top$, $\mathbf q_{1,i} \triangleq -\beta b_i\,\mathbf g_{\alpha,i}^*$, $\mathbf q_{2,i}^\top \triangleq -\alpha^* b_i^*\,\mathbf g_{\beta,i}^\top$, and $t_i \triangleq \alpha^*\beta\,|b_i|^2$.

Combining the quadratic objective \eqref{eq:gain_quad} with the $N_\mathrm S$ quadratic constraints
\eqref{eq:bin_qcqp} yields the following quadratically constrained quadratic program (QCQP):
\begin{equation}
\begin{aligned}
\max_{\mathbf x\in\mathbb C^{N_\mathrm S}}\quad
& \mathbf{x}^\dagger\,\mathbf{R}_0\,\mathbf{x}
+ 2\,\Re\!\left\{\mathbf{q}_0^\top\,\mathbf{x}\right\}
+ t_0\\
\text{s.t.}\ \ \
& \mathbf x^\dagger \mathbf R_i \mathbf x
+\mathbf x^\dagger \mathbf q_{1,i}
+\mathbf q_{2,i}^\top \mathbf x
+t_i = 0,\qquad i=1,\dots,N_\mathrm S .
\end{aligned}
\label{eq:qcqp}
\end{equation}
The QCQP in \eqref{eq:qcqp} is non-convex because it maximizes a convex quadratic objective and it includes quadratic equality constraints. Now, we rewrite these terms such that they are linear in a higher-dimensional variable
\begin{equation}
\mathbf X \triangleq \mathbf x\mathbf x^\dagger \in \mathbb C^{N_\mathrm S\times N_\mathrm S},
\label{eq:X_def}
\end{equation}
where it is obvious upon inspection that $\mathbf{X}$ must have rank one and be positive semidefinite. Specifically, we have
\begin{equation}
        \mathbf x^\dagger \mathbf R_i \mathbf x = \mathrm{tr} \left( \mathbf{R}_i \mathbf{x} \mathbf{x}^\dagger   \right) = \mathrm{tr} \left( \mathbf{R}_i \mathbf{X}    \right)
\end{equation}
for $0 \leq i \leq N_\mathrm{S}$. Now, we can equivalently rewrite the QCQP from (\ref{eq:qcqp}) as
\begin{equation}
\begin{aligned}
\max_{\mathbf x,\mathbf X}\quad
& \mathrm{tr}(\mathbf R_0\mathbf X)
+2\,\Re\!\left\{\mathbf q_0^\top\mathbf x\right\}
+t_0\\
\text{s.t.}\quad
& \mathrm{tr}(\mathbf R_i\mathbf X)
+\mathbf x^\dagger \mathbf q_{1,i}
+\mathbf q_{2,i}^\top \mathbf x
+t_i = 0,\qquad i=1,\dots,N_\mathrm S,\\
& \mathbf X=\mathbf x\mathbf x^\dagger.
\end{aligned}
\label{eq:qcqp_lifted}
\end{equation}
The formulation of our QCQP in \eqref{eq:qcqp_lifted} is still non-convex due to the rank-one constraint implied by
$\mathbf X=\mathbf x\mathbf x^\dagger$.

Since the rank-one constraint is the only non-convex part of the lifted problem, we can obtain a convex semidefinite program (SDP~\cite{vandenberghe1996semidefinite}) by dropping the rank-one constraint. This is the semidefinite relaxation (SDR) of our QCQP~\cite{luo2010semidefinite}. Specifically, we relax the requirement $\mathbf X=\mathbf x\mathbf x^\dagger$ to the requirement $\mathbf X \succeq \mathbf x\mathbf x^\dagger$. To express this requirement as a linear matrix inequality, let us consider the block matrix
\begin{equation}
\mathbf M \triangleq 
\begin{bmatrix}
\mathbf X & \mathbf x\\
\mathbf x^\dagger & 1
\end{bmatrix}.
\label{eq:schur_block}
\end{equation}
Since the bottom-right block is positive ($1>0$), the Schur-complement identity yields the equivalence
\begin{equation}
\mathbf M \succeq \mathbf 0
\quad \Longleftrightarrow \quad
\mathbf X-\mathbf x\mathbf x^\dagger \succeq \mathbf 0,
\label{eq:schur_equiv}
\end{equation}
i.e., \(\mathbf M\succeq \mathbf 0\) is a linear matrix inequality representing $\mathbf X \succeq \mathbf x\mathbf x^\dagger$. The SDR of our QCQP thus becomes the following SDP:
\begin{equation}
\begin{aligned}
\max_{\mathbf x,\mathbf X}\quad
& \mathrm{tr}(\mathbf R_0\mathbf X)
+2\,\Re\!\left\{\mathbf q_0^\top\mathbf x\right\}
+t_0\\
\text{s.t.}\quad
& \mathrm{tr}(\mathbf R_i\mathbf X)
+\mathbf x^\dagger \mathbf q_{1,i}
+\mathbf q_{2,i}^\top \mathbf x
+t_i = 0,\qquad i=1,\dots,N_\mathrm S,\\
& 
\begin{bmatrix}
\mathbf X & \mathbf x\\
\mathbf x^\dagger & 1
\end{bmatrix}\succeq \mathbf 0.
\end{aligned}
\label{eq:sdr_sdp}
\end{equation}
Because any feasible point of the original QCQP \eqref{eq:qcqp_lifted} (with $\mathbf X=\mathbf x\mathbf x^\dagger$) is also feasible for \eqref{eq:sdr_sdp}, the optimal value of \eqref{eq:sdr_sdp} upper-bounds the optimal value of \eqref{eq:qcqp}. The SDP can be solved with standard convex optimization frameworks; we use the CVX modeling framework \cite{cvx} with the SDPT3 (v4.0) solver. 

Our SDR-based bound is defined as the optimal objective value of the relaxed SDP. Given a solver output $(\check{\mathbf{X}},\check{\mathbf{x}})$ as solution of~\eqref{eq:sdr_sdp}, then
\begin{equation}
B_\mathrm{SDR} \triangleq \mathrm{tr}(\mathbf R_0 \,\check{\mathbf X}) +2\,\Re\!\left\{\mathbf q_0^\top\,\check{\mathbf x}\right\} +t_0.
\label{eq:Bsdr_def_star}
\end{equation}

To demonstrate that $B_{\mathrm{SDR}}$ is preserved under the three gauge transformations in Sec.~\ref{sec_Ambiguities}, we show in the Appendix that (i) each gauge induces an invertible change of variables that \textit{bijectively} maps the feasible set of \eqref{eq:sdr_sdp} with parameters $\boldsymbol{\theta}$ onto the feasible set of \eqref{eq:sdr_sdp} with parameters $\tilde{\boldsymbol{\theta}}$, and (ii) this change of variables leaves the SDP objective value unchanged.

While the optimal SDR value $B_{\mathrm{SDR}}$ is gauge-invariant, the eigenvalue spectrum of $\check{\mathbf X}$ is generally \textit{not} gauge-invariant. Hence, the effective rank~\cite{roy2007effective} of $\check{\mathbf X}$, denoted by $R_\mathrm{eff}(\check{\mathbf X})$, and related metrics are generally gauge-dependent and should not be interpreted as an intrinsic indicator of non-convexity or tightness.

\section{Discrete RIS Optimization Algorithms for \\ SISO Channel Gain Maximization}
\label{sec_Opti}

In this section, we briefly describe four algorithms that we subsequently use to optimize a 1-bit-programmable RIS for SISO channel gain enhancement. To be clear, this paper does not claim to propose new algorithms for RIS optimization. Instead, we use these optimization algorithms to see how closely their optimized RIS configurations approach the computed upper bounds, in order to probe the bounds' tightness. 

\textit{Exhaustive Search (ES);}
An exhaustive search (ES) identifies the globally optimal RIS configuration by evaluating the SISO channel gain for each of the $2^{N_{\mathrm S}}$ possible binary control vectors $\mathbf v$. ES is only feasible for sufficiently small $N_\mathrm{S}$. When ES is feasible, ES is valuable because it identifies the global optimum, unlike other optimization algorithms which only tend to identify local optima. Knowing the global optimum allows us to assess the tightness of our bounds. In addition, the performance of other optimization algorithms can be benchmarked against the global optimum. To alleviate the computational burden of ES, we precompute the baseline solution for the $\mathbf{v}=\mathbf{0}$ RIS configuration and then evaluate each bit pattern via a Woodbury update~\cite{prod2023efficient}, which only requires solving a smaller $k\times k$ linear system where $k$ is the number of entries of $\mathbf{v}$ set to one. 

\textit{Coordinate Descent (CD):}
We begin by evaluating the SISO channel gain for $K=100$ random binary RIS configurations, retaining the one with the largest channel gain as initialization for our coordinate descent (CD). During CD, we loop over each RIS element in turn, flip its state, evaluate the new channel gain, and update our currently best RIS configuration if the flip improved the channel gain. Due to the optimization problem's non-convexity, we loop multiple times over the RIS elements until convergence (no change in the currently best RIS configuration for $N_\mathrm{S}$ consecutive iterations). We alleviate the computational cost of evaluating the end-to-end channel after a single-flip change of the RIS configuration with the Woodbury identity~\cite{prod2023efficient}.

\textit{Genetic Algorithm (GA):}
We define the fitness function optimized by the genetic algorithm (GA) as the negative of the SISO channel gain. We use default parameters for our GA. Specifically, we use a population size of 200, and a maximum number of generations of $100 N_\mathrm{S}$. Moreover, we stop early if there was no improvement for the last 50 generations or if the improvement drops below a threshold of $10^{-6}$.

\textit{Projected SDR Solution (P-SDR):}
While the SDR in Sec.~\ref{sec:sdr_bound} yields an upper bound, its optimizer $(\check{\mathbf X},\check{\mathbf x})$ generally does not satisfy $\check{\mathbf X}  = \check{\mathbf x}\,\check{\mathbf x}^\dagger$ and therefore does not directly correspond to a feasible binary RIS configuration. To obtain a feasible binary control vector from the SDR optimizer, we first determine the per-element relaxed reflection-coefficient estimate $\check\rho_i \triangleq \check x_i / \bigl(b_i+(\mathbf\Gamma\check{\mathbf x})_i\bigr)$. Then, we quantize $\check\rho_i$ to the nearest discrete value in $\{\alpha,\beta\}$ (in terms of the Euclidean distance in the complex plane), yielding a binary vector $\check{\mathbf v} \in \mathbb{B}^{N_\mathrm{S}}$ by setting $\check v_i=0$ if the quantized value is $\alpha$ and $\check v_i=1$ if it is $\beta$.

\section{Full-Wave Numerical Results}
\label{sec_Numerical}

In this section, we evaluate the three bounds derived in Sec.~\ref{sec_Bounds} for a numerically simulated SISO channel parametrized by 64 1-bit-programmable RIS elements. Our full-wave numerical simulation provides access to the ground-truth MNT parameters (free of any ambiguity) by replacing the tunable lumped elements with tunable lumped ports~\cite{tapie2023systematic,zheng2024mutual}. In addition, we can freely choose the encoding parameters, including a case in which the loads' reflection coefficients have unit magnitude so that the IBD bound applies. Because the model parameters are not unambiguously known in experimental settings, and the load characteristics are constrained by available components in experiments, these full-wave numerical results provide an added value. 

\begin{figure}
    \centering
    \includegraphics[width=\columnwidth]{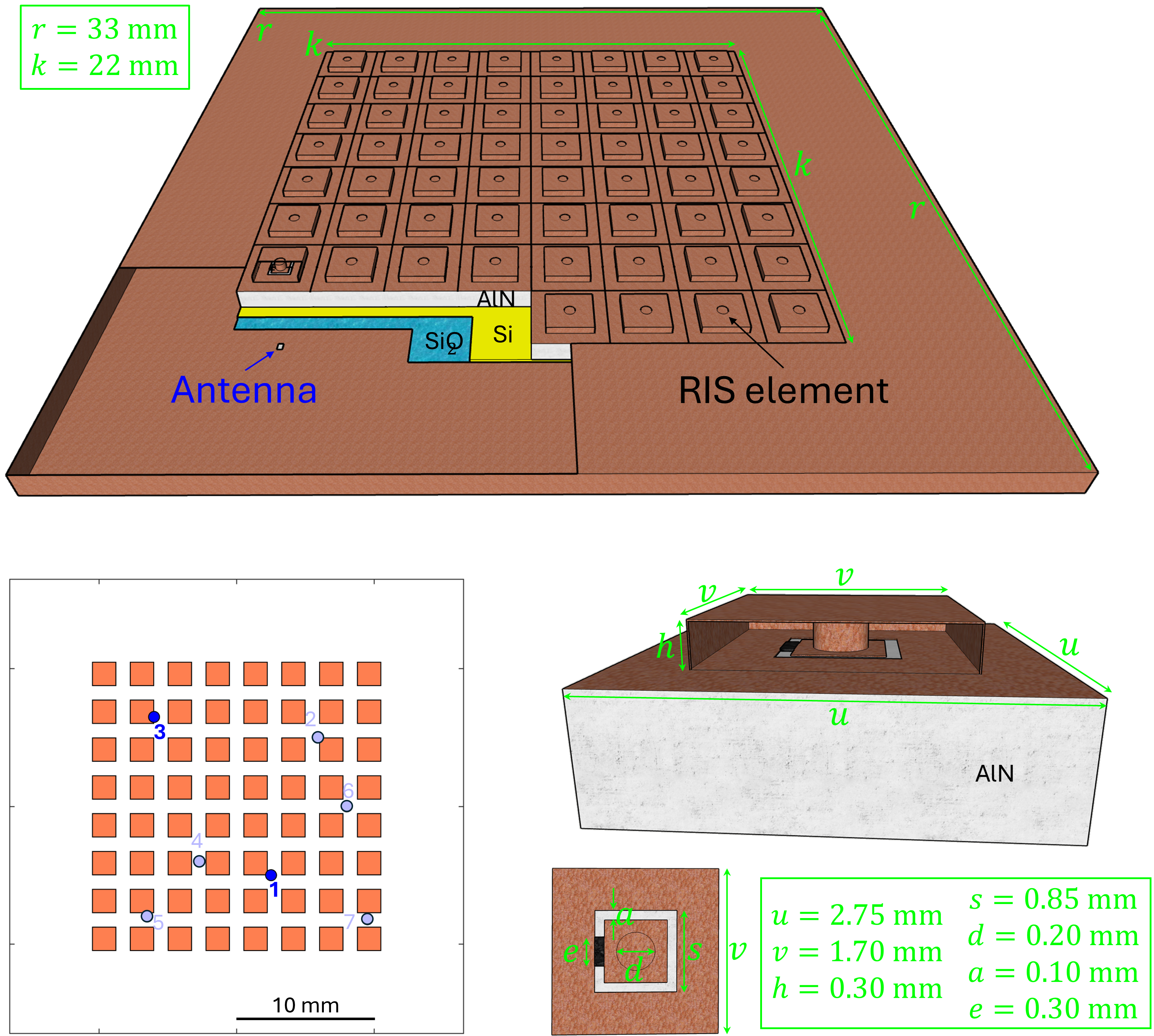}
    \caption{Numerical setup from~\cite{tapie2023systematic} used in Sec.~\ref{sec_Numerical}. The setup comprises 7 antennas (of which those indexed 1 and 3 are used to define the considered SISO channel while the remaining 5 are terminated in matched loads) and 64 RIS elements.}
    \label{Fig1}
\end{figure}

Our full-wave numerical setup emulates the RIS-parametrized wireless network-on-chip (WNoC) described in~\cite{tapie2023systematic}, comprising 64 RIS elements operating in the 60~GHz regime. Although the setup involves 7 antennas, we only consider one pair of antennas to define our SISO channel and assume the remaining 5 antennas are terminated in matched loads. To be clear, our contribution in this section is the evaluation of the bounds from Sec.~\ref{sec_Bounds} in this numerical setup (as well as checking their tightness based on how closely the optimizations from Sec.~\ref{sec_Opti} reach the bounds), but not the design and simulation of the setup itself (which were already reported in~\cite{tapie2023systematic}). Thus, we do not elaborate technical details of the setup's design and simulation here; for completeness, we display relevant details in Fig.~\ref{Fig1} and refer interested readers to~\cite{tapie2023systematic} for further details. In this section, we consider an operating frequency of 60~GHz and select the antennas indexed 1 and 3 seen in Fig.~\ref{Fig1} to define the SISO channel. We consider three different sets of $\{\alpha,\beta\}$:
\begin{itemize}
    \item \textit{PM}: We assume $\alpha=-1$ and $\beta=1$, corresponding to short-circuit and open-circuit terminations. In this case we can apply the IBD bound.
    \item \textit{PIN}: We deduce $\alpha=0.6366-0.7712\jmath$ and $\beta=-0.8116$ from the data sheet of a commercial PIN diode~\cite{tapie2023systematic}.  
    \item \textit{01}: We assume $\alpha=0$ and $\beta=1$, corresponding to matched-load and open-circuit terminations.
\end{itemize}
When we work with a value of $N_\mathrm{S}$ below 64, we fix the termination of the remaining $64-N_\mathrm{S}$ RIS elements to $\alpha$.\footnote{We split the set $\mathcal S$ (ports associated with all RIS elements) into the set $\mathcal S_1$ ($N_\mathrm{S}$ ports associated with the utilized RIS elements) and the set $\mathcal S_2$ ($64-N_\mathrm{S}$ ports associated with unused RIS elements), and partition the RIS-related blocks accordingly as
$\mathbf a=\begin{bmatrix}\mathbf a_{\mathcal S_1}\\ \mathbf a_{\mathcal S_2}\end{bmatrix}$,
$\mathbf b=\begin{bmatrix}\mathbf b_{\mathcal S_1}\\ \mathbf b_{\mathcal S_2}\end{bmatrix}$,
and
$\mathbf\Gamma=\begin{bmatrix}\mathbf\Gamma_{\mathcal S_1\mathcal S_1}&\mathbf\Gamma_{\mathcal S_1\mathcal S_2}\\ \mathbf\Gamma_{\mathcal S_2\mathcal S_1}&\mathbf\Gamma_{\mathcal S_2\mathcal S_2}\end{bmatrix}$.
The unused loads being fixed to $\alpha$ corresponds to $\mathbf\Phi_{\mathcal S_2}=\alpha\,\mathbf I$ and yields the effective reduced-$N_\mathrm S$ model
\begin{align*}
h(\mathbf r_{\mathcal S_1})=h_0'+\mathbf a_{\mathcal S_1}^{\prime\top}\,\bigl(\mathbf I-\mathbf\Phi_{\mathcal S_1}(\mathbf r_{\mathcal S_1})\,\mathbf\Gamma_{\mathcal S_1\mathcal S_1}'\bigr)^{-1}\,\mathbf\Phi_{\mathcal S_1}(\mathbf r_{\mathcal S_1})\,\mathbf b_{\mathcal S_1}',
\end{align*}
where
\begin{align*}
\mathbf\Gamma_{\mathcal S_1\mathcal S_1}'&=\mathbf\Gamma_{\mathcal S_1\mathcal S_1}+\mathbf\Gamma_{\mathcal S_1\mathcal S_2}\bigl(\mathbf I-\mathbf\Phi_{\mathcal S_2}\mathbf\Gamma_{\mathcal S_2\mathcal S_2}\bigr)^{-1}\mathbf\Phi_{\mathcal S_2}\mathbf\Gamma_{\mathcal S_2\mathcal S_1},\\
\mathbf b_{\mathcal S_1}'&=\mathbf b_{\mathcal S_1}+\mathbf\Gamma_{\mathcal S_1\mathcal S_2}\bigl(\mathbf I-\mathbf\Phi_{\mathcal S_2}\mathbf\Gamma_{\mathcal S_2\mathcal S_2}\bigr)^{-1}\mathbf\Phi_{\mathcal S_2}\mathbf b_{\mathcal S_2},\\
\mathbf a_{\mathcal S_1}^{\prime\top}&=\mathbf a_{\mathcal S_1}^\top+\mathbf a_{\mathcal S_2}^\top\bigl(\mathbf I-\mathbf\Phi_{\mathcal S_2}\mathbf\Gamma_{\mathcal S_2\mathcal S_2}\bigr)^{-1}\mathbf\Phi_{\mathcal S_2}\mathbf\Gamma_{\mathcal S_2\mathcal S_1},\\
h_0'&=h_0+\mathbf a_{\mathcal S_2}^\top\bigl(\mathbf I-\mathbf\Phi_{\mathcal S_2}\mathbf\Gamma_{\mathcal S_2\mathcal S_2}\bigr)^{-1}\mathbf\Phi_{\mathcal S_2}\mathbf b_{\mathcal S_2}.
\end{align*}
}

\begin{figure*}
    \centering
    \includegraphics[width=2\columnwidth]{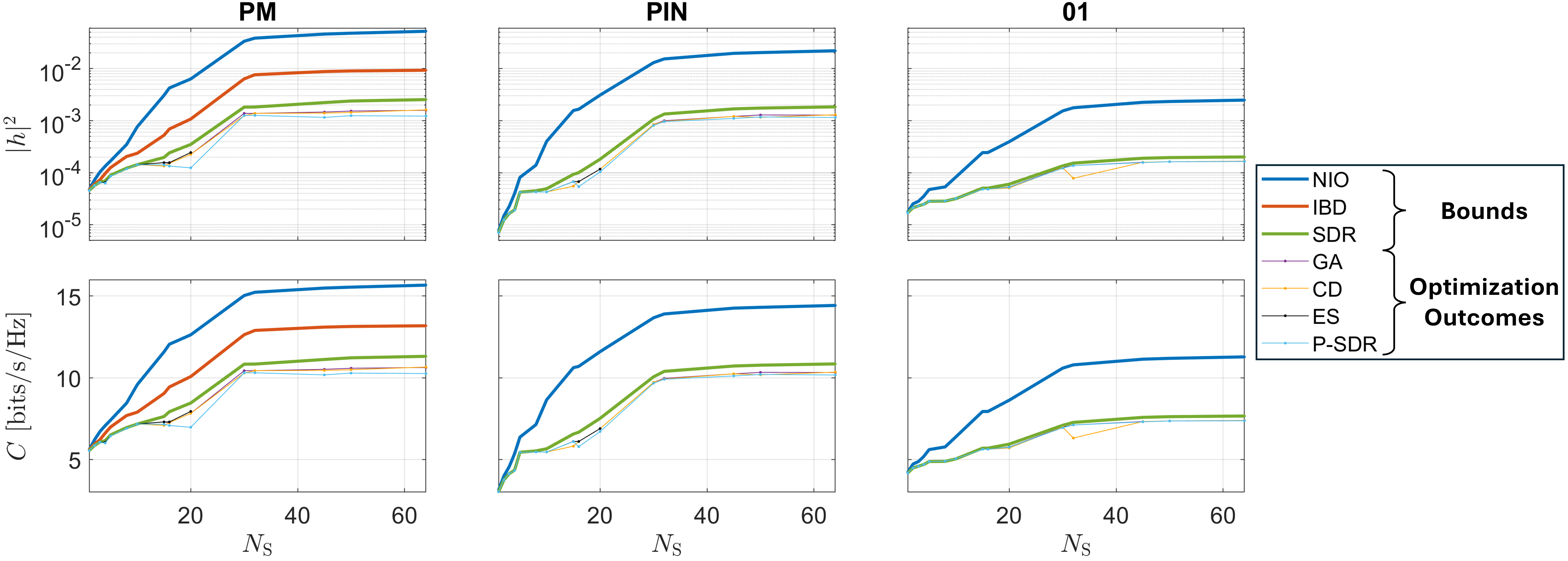}
    \caption{Bounds (NIO, IBD, SDR -- see Sec.~\ref{sec_Bounds}) on the achievable channel gain (top row) and the Shannon capacity (with $P_\mathrm{T}=10\ \mathrm{mW}$ and $\sigma^2 = 10^{-5}\ \mathrm{mW}$; bottom row) in the RIS-parametrized SISO setup in Fig.~\ref{Fig1} (all but two antennas are terminated in matched loads) as a function of $N_\mathrm{S}$ (horizontal axis) and for three different choices of $\{\alpha,\beta\}$ (different panels: PM, PIN, 01). The displayed bounds are the ones derived in Sec.~\ref{sec_Bounds}; the IBD bound is only shown for the PM case because it requires the two available loads to have unit magnitude. In addition to the bounds, the outcomes of the four considered optimization strategies (GA, CD, ES, P-SDR -- see Sec.~\ref{sec_Opti}) are displayed; ES is only shown for $N_\mathrm{S}\leq 20$. }
    \label{Fig2}
\end{figure*}

Our results are displayed in Fig.~\ref{Fig2} and reveal a clear and consistent hierarchy between the bounds: the SDR bound is systematically tighter than the IBD bound (when it exists), which in turn is systematically tighter than the NIO bound. In fact, our results show that the SDR bound is very tight. Up to $N_\mathrm{S}=10$, our optimization techniques exactly reach the SDR bound for the three cases of different load choices. Even for $N_\mathrm{S}=64$, our SDR bound appears to be quite tight. While we do not know the globally optimal SISO channel gain since ES is not feasible for $N_\mathrm{S}=64$, we can still compare the bound to the channel gain $|h_\mathrm{opt}|^2$ achieved with the best of our optimized RIS configurations. As summarized in Table~\ref{tab:bound_tightness_numerical}, the ratio of our SDR bound and $|h_\mathrm{opt}|^2$ is very small, being 1.57 in the worst case (PM) and 1.21 in the best case (01). In contrast, the ratio is 5.76 in the PM case (the IBD bound cannot be evaluated for the other cases), and between 32.23 (PM) and 14.93 (01) for the NIO bound. The SDR bound is thus more than an order of magnitude tighter than the NIO bound, and 3.66 times tighter than the IBD bound in the case in which the IBD bound applied. While this hierarchy among the three bounds was expected given their assumptions and account of real-world constraints, the tightness of the SDR bound is remarkable. One observation based on Fig.~\ref{Fig2} and Table~\ref{tab:bound_tightness_numerical} is that the bound tightness varies slightly with the available loads; specifically, it seems that the NIO and SDR bounds are tighter when the loads are lossier. 

The dependence of the bounds on the choice of $\{\alpha,\beta\}$ is plausible because the attenuation in the loads affects the weight with which multi-bounce paths between RIS elements contribute to the end-to-end channel. The dependence of the bounds on the number of RIS elements also makes sense because the binary search space scales exponentially with $N_\mathrm{S}$; moreover, the number of significant multi-bounce paths between RIS elements increases with $N_\mathrm{S}$, too.

The second row of Fig.~\ref{Fig2} shows that all observations for the SISO channel gain carry over to the Shannon capacity, as expected given the monotone mapping between the two.

As a side remark, we observe in Fig.~\ref{Fig2} that GA and CD usually perform comparably well (in line with related theoretical results in~\cite{hammami2025statistical}) and actually reach at least two thirds of the tightest bound. The P-SDR result is comparable to the GA and CD result in most cases, but slightly inferior in particular in the PM case. Another interesting side observation in Fig.~\ref{Fig2} is that the achievable SISO channel gain saturates for $N_\mathrm{S}>32$, indicating diminishing marginal returns of adding more RIS elements beyond $N_\mathrm{S}=32$.

\begin{table}
\caption{NIO, IBD, and SDR bounds relative to the best optimized SISO channel gain $|h_\mathrm{opt}|^2$, for the setup in Fig.~\ref{Fig1} with $N_\mathrm{S}=64$. }
\label{tab:bound_tightness_numerical}
\centering
\renewcommand{\arraystretch}{1.15}
\begin{tabular}{lcccc}
\hline
 & $B_\mathrm{NIO}/|h_\mathrm{opt}|^2$ & $B_\mathrm{IBD}/|h_\mathrm{opt}|^2$ & $B_\mathrm{SDR}/|h_\mathrm{opt}|^2$ & $R_\mathrm{eff}(\check{\mathbf X})$ \\

\hline
PM & 32.23 & 5.76 & 1.57 & 2.05 \\
PIN & 16.96 & N/A & 1.42 & 1.69 \\
01 & 14.93 & N/A & 1.21 & 1.69 \\
\hline
\end{tabular}
\end{table}

\section{Experimental Results}
\label{sec_Experimental}

\begin{figure*}
    \centering
    \includegraphics[width=2\columnwidth]{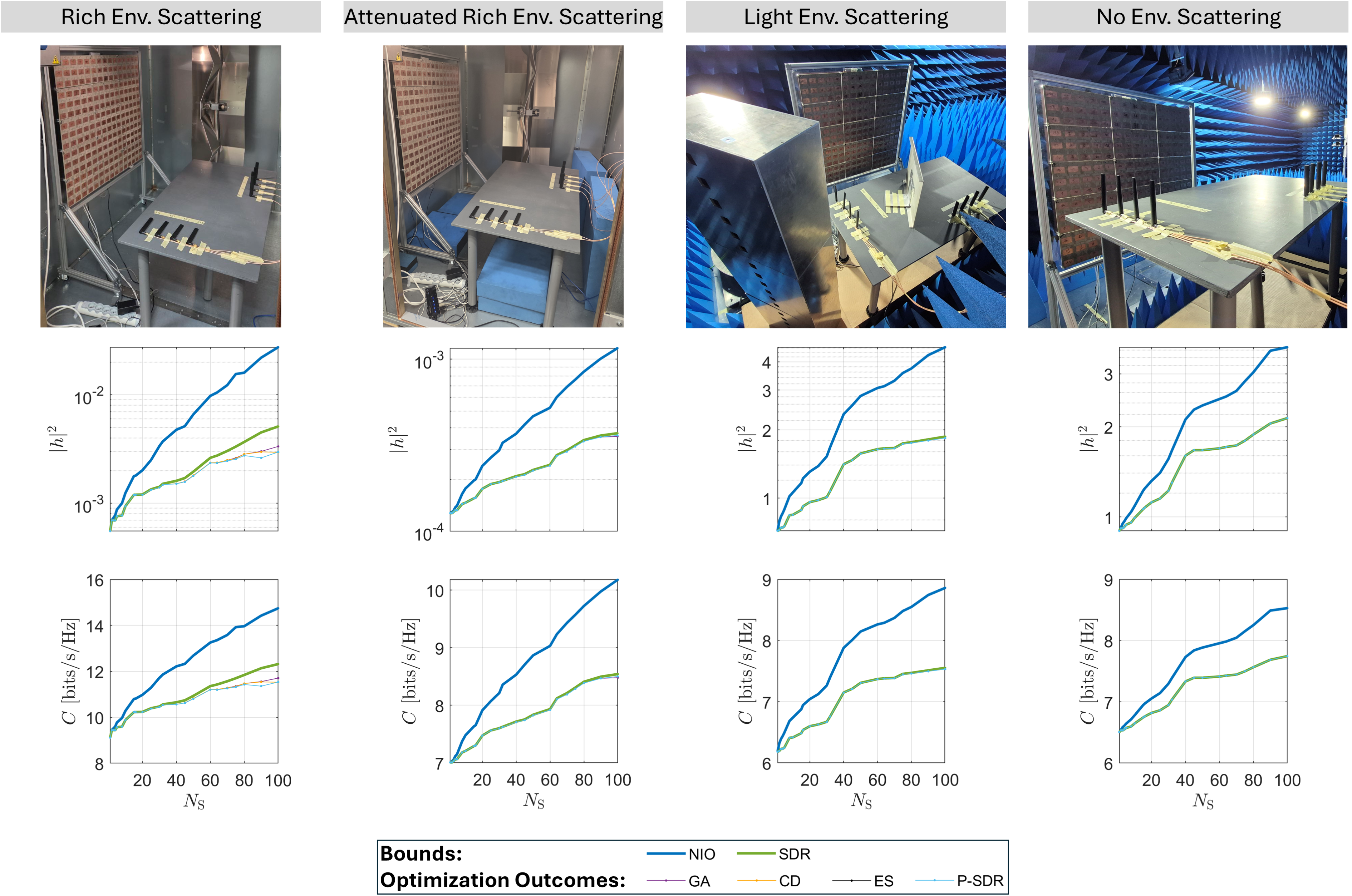}
    \caption{Bounds (NIO, SDR -- see Sec.~\ref{sec_Bounds}) on the achievable channel gain (middle row) and the Shannon capacity (with $P_\mathrm{T}=10\ \mathrm{mW}$ and $\sigma^2 = 10^{-5}\ \mathrm{mW}$; bottom row) in the RIS-parametrized SISO setups displayed in the top row (all but two antennas are terminated in matched loads), as a function of $N_\mathrm{S}$ (horizontal axis). In addition to the bounds, the outcomes of the four considered optimization strategies (GA, CD, ES, P-SDR -- see Sec.~\ref{sec_Opti}) are displayed; ES is only shown for $N_\mathrm{S}\leq 20$. }
    \label{Fig3}
\end{figure*}

In this section, we evaluate the NIO and SDR bounds derived in Sec.~\ref{sec_Bounds} for four distinct real-world experimental SISO channels parametrized by 100 1-bit-programmable RIS elements; the IBD bound is not applicable because the loads are not lossless. The four experiments are based on the same RIS prototype but differ regarding the environmental scattering, which in turn influences the mutual coupling between the RIS elements, as seen in the experiments reported in~\cite{rabault2023tacit,del2025experimental}. Our RIS prototype is composed of 225 half-wavelength-sized RIS elements designed for operation around 2.45~GHz of which we only use 100 RIS elements (the remaining ones are always in a fixed reference configuration). Each RIS element comprises a PIN diode which is the tunable lumped element described as ``virtual'' port terminated by a tunable load in our system model. The description of these PIN diodes  as tunable lumped elements applies because  the PIN diode is much smaller than the wavelength and was validated in~\cite{del2025experimental,del2025frozen}. Additional details on the RIS design are available in~\cite{ahmed2025over}.
Our setup involves 4 transmitting and 4 receiving antennas, but we only use one transmitting and one receiving antenna to define the SISO channel in each case; the remaining six antennas are terminated with matched loads. Because our system model concerns the overall end-to-end channel, it remains applicable regardless of how the antennas are oriented and irrespective of the surrounding propagation environment. Throughout this section, we consider the end-to-end SISO channel at 2.45~GHz.

As already explained in Sec.~\ref{sec_Ambiguities}, the model parameters required to describe any of our four experimental settings are neither known in closed form, nor can they be obtained from numerical simulations, nor can they be measured directly in the experiment. However, a proxy set of model parameters $\tilde{\boldsymbol{\theta}}$ can be identified based on measurements of the end-to-end channel for a series of known RIS control vectors~\cite{sol2024experimentally,del2025physics,ContRIS_LWC,largeRIS_TCOM,del2025experimental}. Here, we use for each experimental setup a set of proxy parameters estimated with the method described in~\cite{ContRIS_LWC} based on measurements of the $4 \times 4$ end-to-end channel using an eight-port VNA (two cascaded Keysight P5024B 4-port VNAs); for our analysis in the present work, we select the SISO channel of interest and assume the remaining six antennas are terminated with matched loads. 

We display the four radio environments in our experiments in the top row of Fig.~\ref{Fig3}; their characteristics are the following:
\begin{itemize}
    \item \textit{Rich Env. Scattering}: Reverberation chamber (RC, $1.75\ \mathrm{m}\times1.5\ \mathrm{m}\times2\ \mathrm{m}$).
    \item \textit{Attenuated Rich Env. Scattering}: Same RC as before but loaded with absorbing material. 
    \item \textit{Light Env. Scattering}: Anechoic chamber (AC) containing a few metallic scattering objects. 
    \item \textit{No Env. Scattering}: AC without any additional scattering objects. 
\end{itemize}
The RC has a mode stirrer, as seen in Fig.~\ref{Fig3}, but that stirrer is static throughout all of our experiments. 

\begin{table}
\caption{NIO and SDR bounds relative to the best optimized SISO channel gain $|h_\mathrm{opt}|^2$, as well as effective rank of the SDR optimizer, for the four setups in Fig.~\ref{Fig3} with $N_\mathrm{S}=100$.}
\label{tab:bound_tightness_exp}
\centering
\renewcommand{\arraystretch}{1.15}
\begin{tabular}{lccc}
\hline
Env. Scatt. & $B_\mathrm{NIO} / |h_\mathrm{opt}|^2$ &  $B_\mathrm{SDR} / |h_\mathrm{opt}|^2$  & $R_\mathrm{eff}(\check{\mathbf X})$\\
\hline
Rich & 8.21 & 1.53 & 1.99 \\
Attenuated Rich & 3.21 & 1.03 & 1.33 \\
Light & 2.51 & 1.01 & 1.43 \\
None & 1.72 & 1.00 & 1.01 \\
\hline
\end{tabular}
\end{table}

Our results are displayed in Fig.~\ref{Fig3}. We observe that the SDR bound is systematically tighter than the NIO bound. Moreover, we (almost) reach the SDR bound with the considered optimization techniques for $N_\mathrm{S}<32$ in the rich-scattering environment, and for all considered $N_\mathrm{S}$ in the other three radio environments. It is plausible that the bound is tighter when mutual coupling between the RIS elements is weaker, which is in line with our observation because rich scattering contributes significantly to the mutual coupling between the RIS elements, as documented in~\cite{rabault2023tacit,del2025experimental}. As in Sec.~\ref{sec_Numerical}, we also observe that the bounds are tighter for smaller $N_\mathrm{S}$. As summarized in Table~\ref{tab:bound_tightness_exp}, the ratio of our SDR bound and the channel gain achieved with the best optimization result is 1.53 for the rich scattering environment with $N_\mathrm{S}=100$. Meanwhile, for the three other radio environments, the ratio is essentially unity, meaning that we (almost) attain the bound and can thus certify that no other optimization strategy will be able to substantially outperform the ones we considered in terms of the achieved channel gain. Moreover, for these three radio environments, all three optimizations (GA, CD, P-SDR) yield the same optimized configuration. Remarkably, $R_\mathrm{eff}(\check{\mathbf X})=1.01$ for the free-space radio environment, meaning that the SDR is essentially exact despite the underlying binary constraints, which in turn implies that the optimized P-SDR configuration is globally optimal in this case. Moreover, comparing the attenuated rich scattering case and the light scattering case, illustrates that $R_\mathrm{eff}(\check{\mathbf X})$ is not a reliable indicator of tightness, as explained earlier. The last row of Fig.~\ref{Fig3} shows that all observations for the SISO channel gain carry over to the Shannon capacity, as expected given the monotone mapping between the two.

\section{Discussion}
\label{sec_discussion}

The practical value of our bounds lies in the fact that, \textit{first}, we are able to evaluate them for real-world experimental systems (despite inevitable ambiguities in experimental estimations of the system model parameters), and, \textit{second}, our bounds appear to be very tight thanks to taking into account both the electromagnetically consistent description of wave propagation via MNT \textit{and} the practical hardware constraints of 1-bit-programmable RIS elements (which are very common in most RIS prototypes).

The generality of our system model in Sec.~\ref{sec_SystemModel} implies that the bounding strategies presented in this work directly apply to any linear wave system parametrized by tunable lumped elements. Beyond RIS-parametrized radio environments, examples include various reconfigurable antenna architectures (dynamic metasurface antennas~\cite{tapie2025experimental,almunif2025network}, electrically steerable passive array radiators (ESPARs)~\cite{harrington1978reactively,kawakami2005electrically}, reconfigurable pixel antennas~\cite{Flaviis_PixelAntenna,rodrigo2012frequency,MURCH_TAP_PixelAntenna}) and wave-domain physical neural networks~\cite{momeni2023backpropagation,pWDCperspective}.

Our bounds also directly apply to \textit{real-life} wave systems with so-called ``beyond-diagonal'' tunability, such as beyond-diagonal RISs (BD-RISs~\cite{shen2021modeling}) and beyond-diagonal DMAs (BD-DMAs~\cite{bddma}). The reason is that a realistic embodiment of the ``beyond-diagonal'' load network terminating the RIS elements relies on an ensemble of $N_\mathrm{C}$ tunable lumped elements, such that a physics-consistent model with the same mathematical structure as the one described in Sec.~\ref{sec_SystemModel} for a conventional RIS applies, as derived in~\cite{del2025physics}. The  approach underlying~\cite{del2025physics} is exactly the same idea that we described in Sec.~\ref{sec_SystemModel}: we partition the system into three entities: (i) antenna ports, (ii) tunable lumped elements, and (iii) all other static components. In the case of a beyond-diagonal RIS with $N_\mathrm{S}$ RIS elements and $N_\mathrm{C}$ tunable lumped elements in the beyond-diagonal load network instead of a conventional RIS with $N_\mathrm{S}$ RIS elements terminated by individual loads, entity (ii) comprises $N_\mathrm{C}$ instead of $N_\mathrm{S}$ tunable lumped elements, and entity (iii) comprises both the structural and environmental scattering as well as static components of the load network. In short, it is possible to treat the connection between radio environment and the static components of the beyond-diagonal load network as an \textit{effective} radio environment with $N_\mathrm{C}$ ``virtual'' ports terminated by \textit{individual} tunable loads. 
Notwithstanding, common idealized assumptions about BD-RIS can be useful, as evidenced by the elegant closed-form global solution underlying our IBD bound in Sec.~\ref{subsec_IBD} based on~\cite{wu2025beyond}. However, the gap to reality consists in the issue of conceiving a practical load circuit capable of being tuned to realize any desired $\mathbf{\Phi}$ that is unitary and symmetric. In practice, the connections between tunable components in the beyond-diagonal load network are generally \textit{not} lossless and delayless, the tunability of the lumped elements is generally subject to severe quantization constraints, and the tunable lumped elements are generally lossy. A system model accounting for these real-life constraints ends up having the same mathematical structure as the system model for conventional RIS used in this work, as argued above. Thus, \textit{real-life} BD-RIS parametrized SISO channel gain can be bounded with the techniques described in this work.

\section{Conclusion}
\label{sec_conclusion}

To summarize, we have derived three electromagnetically consistent bounds on information transfer via a wireless channel parametrized by 1-bit-programmable RIS elements; we evaluated the bounds for one numerical and four experimental scenarios as a function of the number of RIS elements, and we probed the tightness of the bounds by comparing them to the outcomes of multiple techniques for discrete optimization of the RIS elements. The SDR-based bound was very tight in all considered scenarios; our discrete optimizations reached at least 64\ \% (but often 100\ \%) of our SDR-based bound. 

To the best of our knowledge, our present work describes the first electromagnetically consistent bounds on information transfer in real-life programmable channels with realistic feasibility sets for the RIS configurations. Moreover, our work constitutes the first experimental validation of these bounds. Thereby, our work contributes to the development of an electromagnetic information theory for \textit{programmable} channels.

Looking forward, we envision to extend the presented method to information transfer via real-world programmable MIMO channels, as well as to goal-oriented information transfer in real-world programmable channels.

\appendix

\textit{Bijection of feasible sets of the SDP:} Our strategy consists in demonstrating for each of the three gauge types in turn that they result in an invertible mapping $(\mathbf x,\mathbf X)\mapsto(\tilde{\mathbf x},\tilde{\mathbf X})$
that preserves all SDP constraints. 
Applying (\ref{eq:gauge_diag}) to (\ref{eq:bd_x_def}), we see that for the diagonal-similarity gauge $\tilde{\mathbf{x}} = \mathbf{D}\,\mathbf{x}$, and thus $\tilde{\mathbf{X}} = \mathbf{D}\,\mathbf{X}\,\mathbf{D}^\dagger$. Applying (\ref{eq:gauge_scale}) to (\ref{eq:bd_x_def}), we see that for the complex-scaling gauge $\tilde{\mathbf{x}} = c\,\mathbf{x}$, and thus $\tilde{\mathbf{X}} = |c|^2\,\mathbf{X}$. Applying (\ref{eq:gauge_mobius}) to (\ref{eq:bd_x_def}), we see that for the M\"obius gauge $\tilde{\mathbf{x}} = \mathbf{W}\,\mathbf{x} + \mathbf{w}$, where $\mathbf{W}=(\mathbf{I}_{N_\mathrm{S}}-m\,\mathbf{\Gamma})/k$ and $\mathbf{w} = -m \, \mathbf{b} / k$, and thus $\tilde{\mathbf{X}} = \mathbf{W}\,\mathbf{X}\,\mathbf{W}^\dagger + \mathbf{W}\,\mathbf{x}\,\mathbf{w}^\dagger+ \mathbf{w}\,\mathbf{x}^\dagger\,\mathbf{W}^\dagger + \mathbf{w}\,\mathbf{w}^\dagger$. All three gauge types thus transform $\mathbf{M}$ into $\tilde{\mathbf{M}} = \mathbf{T}\,\mathbf{M}\,\mathbf{T}^\dagger$, where $\mathbf T_\mathrm{DS}=\mathrm{blkdiag}(\mathbf D,1)$ for the diagonal-similarity gauge, $\mathbf T_\mathrm{CS}=\mathrm{blkdiag}(c\mathbf I_{N_\mathrm{S}},1)$ for the complex-scaling gauge, and $\mathbf T_\mathrm{MO}=\begin{bmatrix}\mathbf W & \mathbf w\\ \mathbf 0^\top & 1\end{bmatrix}$ for the M\"obius gauge. For admissible gauge parameters, $\mathbf T$ is invertible; hence the mapping $\mathbf M \mapsto \tilde{\mathbf M}=\mathbf T\,\mathbf M\,\mathbf T^\dagger$ is bijective with its inverse being $\tilde{\mathbf M} \mapsto \mathbf M=\mathbf T^{-1}\,\tilde{\mathbf M}\,\mathbf T^{-\dagger}$. Moreover, since $\mathbf T$ is invertible, the transformation $\tilde{\mathbf M}=\mathbf T\,\mathbf M\,\mathbf T^\dagger$ preserves positive semidefiniteness in both directions, i.e., $\mathbf M\succeq\mathbf 0$ if and only if $\tilde{\mathbf M}\succeq\mathbf 0$.

Finally, we show that the constraints of our SDP are preserved. Our strategy consists in rewriting the constraint term in the second line of~(\ref{eq:sdr_sdp}) as
$\mathrm{tr}(\mathbf R_i\mathbf X)
+\mathbf x^\dagger \mathbf q_{1,i}
+\mathbf q_{2,i}^\top \mathbf x + t_i
= \mathrm{tr}\!\big(\mathbf G_i(\boldsymbol\theta)\,\mathbf M\big)$, where
$\mathbf G_i(\boldsymbol\theta)\triangleq
\begin{bmatrix}
\mathbf R_i & \mathbf q_{1,i}\\
\mathbf q_{2,i}^\top & t_i
\end{bmatrix}$.
We then demonstrate for each of the three gauge types that the transformed parameters
$\tilde{\boldsymbol\theta}$ yield a matrix $\mathbf G_i(\tilde{\boldsymbol\theta})$ satisfying
$\mathbf G_i(\tilde{\boldsymbol\theta})
= \eta_i
\left(\mathbf T^{\dagger}\right)^{-1}\,\mathbf G_i(\boldsymbol\theta)\,\mathbf T^{-1}$
with $\eta_i\neq 0$, where the corresponding lifted decision variable transforms as
$\tilde{\mathbf M}=\mathbf T\,\mathbf M\,\mathbf T^\dagger$.
It follows immediately that
$\mathrm{tr}\!\big(\mathbf G_i(\tilde{\boldsymbol\theta})\,\tilde{\mathbf M}\big)
=\eta_i\,\mathrm{tr}\!\big(\mathbf G_i(\boldsymbol\theta)\,\mathbf M\big)$,
and hence $\mathrm{tr}\!\big(\mathbf G_i(\boldsymbol\theta)\,\mathbf M\big)=0$ holds if and only if
$\mathrm{tr}\!\big(\mathbf G_i(\tilde{\boldsymbol\theta})\,\tilde{\mathbf M}\big)=0$.
For the diagonal-similarity gauge, we have $\tilde x_i-\tilde{\rho}\tilde z_i=d_i(x_i-\rho z_i)$ (with $d_i\neq 0$); it follows that $\eta_i^\mathrm{DS}=|d_i|^2$ and we find $\mathbf G_i(\tilde{\boldsymbol\theta}) = \eta_i^\mathrm{DS}\,\left(\mathbf T_\mathrm{DS}^{\dagger}\right)^{-1}\,\mathbf G_i(\boldsymbol\theta)\,\mathbf T_\mathrm{DS}^{-1}$.
For the complex-scaling gauge, we have $\tilde x_i-\tilde\rho\,\tilde z_i=c(x_i-\rho z_i)$ with $c\neq 0$; it follows that $\eta_i^\mathrm{CS}=|c|^2$ and we find $\mathbf G_i(\tilde{\boldsymbol\theta}) = \eta_i^\mathrm{CS}\,\left(\mathbf T_\mathrm{CS}^{\dagger}\right)^{-1}\,\mathbf G_i(\boldsymbol\theta)\,\mathbf T_\mathrm{CS}^{-1}$.
For the M\"obius gauge, we have $\tilde x_i-\tilde\rho\,\tilde z_i=\frac{k}{1-m^\ast\rho}\,(x_i-\rho z_i)$ with $1-m^\ast\rho\neq 0$; it follows that $\eta_i^\mathrm{MO}=\left(\frac{k}{1-m^\ast \alpha}\right)^\ast \left(\frac{k}{1-m^\ast \beta}\right)$ and we find $\mathbf G_i(\tilde{\boldsymbol\theta}) = \eta_i^\mathrm{MO}\,\left(\mathbf T_\mathrm{MO}^{\dagger}\right)^{-1}\,\mathbf G_i(\boldsymbol\theta)\,\mathbf T_\mathrm{MO}^{-1}$.

\textit{Preservation of SDP objective:}
Our strategy consists in rewriting the objective term in the first line of (\ref{eq:sdr_sdp}) as
$\mathrm{tr}(\mathbf R_0\mathbf X) + 2\Re\{\mathbf q_0^\top\mathbf x\}+t_0
= \mathrm{tr}\!\big(\mathbf G_0(\boldsymbol\theta)\,\mathbf M\big)$, where
$\mathbf G_0(\boldsymbol\theta)\triangleq
\begin{bmatrix}
\mathbf R_0 & \mathbf q_0^\ast\\
\mathbf q_0^\top & t_0
\end{bmatrix}$, and then demonstrating for each of the three gauge types that the transformed parameters
$\tilde{\boldsymbol\theta}$ yield a matrix $\mathbf G_0(\tilde{\boldsymbol\theta})$ satisfying
$\mathbf G_0(\tilde{\boldsymbol\theta})
=
\left(\mathbf T^{\dagger}\right)^{-1}\,\mathbf G_0(\boldsymbol\theta)\,\mathbf T^{-1}$.
Then, the preservation of the SDP objective follows immediately:
$\mathrm{tr}\!\big(\mathbf G_0(\tilde{\boldsymbol\theta})\,\tilde{\mathbf M}\big)
=
\mathrm{tr}\!\left(\mathbf T^{-\dagger}\mathbf G_0(\boldsymbol\theta)\mathbf T^{-1}\,
\mathbf T\mathbf M\mathbf T^\dagger\right)
=
\mathrm{tr}\!\big(\mathbf G_0(\boldsymbol\theta)\,\mathbf M\big)$.
Applying (\ref{eq:gauge_diag}) to (\ref{eq:gain_quad}), we see that for the diagonal-similarity gauge $\tilde{\mathbf R}_0 =\mathbf D^{-\dagger}\,\mathbf R_0\,\mathbf D^{-1}$, $\tilde{\mathbf q}_0^\top = \mathbf q_0^\top\,\mathbf D^{-1}$, and $\tilde t_0=t_0$, implying $\mathbf G_0(\tilde{\boldsymbol\theta}) = \left(\mathbf T_\mathrm{DS}^{\dagger}\right)^{-1}\,\mathbf G_0(\boldsymbol\theta)\,\mathbf T_\mathrm{DS}^{-1}$.
Applying (\ref{eq:gauge_scale}) to (\ref{eq:gain_quad}), we see that for the complex-scaling gauge $\tilde{\mathbf R}_0 =\frac{1}{|c|^2}\mathbf R_0$, $\tilde{\mathbf q}_0^\top =\frac{1}{c}\mathbf q_0^\top$, and $\tilde t_0=t_0$, implying $\mathbf G_0(\tilde{\boldsymbol\theta}) = \left(\mathbf T_\mathrm{CS}^{\dagger}\right)^{-1}\,\mathbf G_0(\boldsymbol\theta)\,\mathbf T_\mathrm{CS}^{-1}$.
Applying (\ref{eq:gauge_mobius}) to (\ref{eq:gain_quad}), we see that for the M\"obius gauge
$\tilde{\mathbf R}_0 =\mathbf W^{-\dagger}\,\mathbf R_0\,\mathbf W^{-1}$, $
\tilde{\mathbf q}_0^\top
= \mathbf q_0^\top\,\mathbf W^{-1}-\mathbf w^\dagger\,\mathbf W^{-\dagger}\,\mathbf R_0\,\mathbf W^{-1}$, and $\tilde t_0
= t_0 - 2\,\Re\!\left\{\mathbf q_0^\top\,\mathbf W^{-1}\mathbf w\right\}
+ \mathbf w^\dagger\,\mathbf W^{-\dagger}\,\mathbf R_0\,\mathbf W^{-1}\mathbf w$,
which implies
$\mathbf G_0(\tilde{\boldsymbol\theta}) = \left(\mathbf T_\mathrm{MO}^{\dagger}\right)^{-1}\,\mathbf G_0(\boldsymbol\theta)\,\mathbf T_\mathrm{MO}^{-1}$.

\section*{Acknowledgment}
P.d.H. acknowledges stimulating discussions with O.~D.~Miller and S.~Molesky. P.d.H. further acknowledges J.~Tapie as well as I.~Ahmed, F. Boutet, and C. Guitton, who, under P.d.H.'s supervision, previously conducted the WNoC simulation for the work presented in~\cite{tapie2023systematic} and built the RIS prototype for the work presented in~\cite{ahmed2025over}, respectively. Moreover, P.d.H. acknowledges J.~Sol, who provided technical support for setting up the experiments.  
The authors further acknowledge IETR's QOSC test facility (which is part of the CNRS RF-Net network) and the MIDAS infrastructure of the Aalto School of Electrical Engineering.

\bibliographystyle{IEEEtran}
%\bibliography{references}

\providecommand{\noopsort}[1]{}\providecommand{\singleletter}[1]{#1}%

\end{document}